\documentclass{article}
\usepackage[utf8]{inputenc}
\usepackage{amssymb}
\usepackage{amsmath}
\usepackage{amsthm}
\usepackage{amsfonts}
\usepackage{amssymb}
\usepackage{siunitx}
\usepackage{calc}
\usepackage{mathtools}
\usepackage{latexsym}
\usepackage{accents}
\usepackage{caption}
\usepackage{subcaption}
\usepackage[margin=3cm]{geometry}
\usepackage{booktabs,multicol,multirow,tabularx,array}  
\usepackage{algorithm}
\usepackage{algpseudocode}
\usepackage{diagbox}
\usepackage[pdftex]{graphicx}
\usepackage[dvipsnames]{xcolor}
\usepackage{url}

\newtheorem{remark}{Remark}
\newtheorem{theorem}{Theorem}

\newcommand{\R}{\mathbb{R}}
\newcommand{\I}{\mathcal{I}}
\newcommand{\OO}{\mathcal{O}}

\newcommand{\openOmega}{\accentset{\circ}{\Omega}}

\title{Boundary-safe PINNs extension: Application to non-linear parabolic PDEs in counterparty credit risk\footnote{The authors have no conflict of interest to disclose.}}

\author{Joel P. Villarino$^{1, 2}$ \ and \ Álvaro Leitao$^{1,2}$ \ and \ José A. García-Rodríguez$^{1,2}$}
\date{\today}

\begin{document}

    \maketitle
    \vspace{-0.5cm}
    \begin{center}
        {\footnotesize
        $1$ M2NICA research group, University of Coruña, Spain \\
        $2$ CITIC research center, Spain \\
        E-mails: joel.perez.villarino@udc.es / alvaro.leitao@udc.es / jose.garcia.rodriguez@udc.es
        }
    \end{center}
    
    \bigskip
    \noindent
    
    {\small{\bf ABSTRACT.}
    
        The goal of this work is to develop deep learning numerical methods for solving option XVA pricing problems given by non-linear PDE models. A novel strategy for the treatment of the boundary conditions is proposed, which allows to get rid of the heuristic choice of the weights for the different addends that appear in the loss function related to the training process. It is based on defining the losses associated to the boundaries by means of the PDEs that arise from substituting the related conditions into the model equation itself. Further, automatic differentiation is employed to obtain accurate approximation of the partial derivatives.
    }
    
    \medskip
    \noindent
    {\small{\bf Keywords} : deep learning, PDEs, PINNs, boundary conditions, nonlinear, automatic differentiation, option pricing, XVA.
    }
    
    \baselineskip=\normalbaselineskip
    \section{Introduction} \label{sec:intro}
    Deep learning techniques are machine learning algorithms based on neural networks, also known as artificial neural networks (ANNs), and representation learning, see \cite{goodfellow} and the references therein. From a mathematical point of view, ANNs can be interpreted as multiple chained compositions of multivariate functions, and deep neural networks is the term used for ANNs with several interconnected layers. Such networks are known for being universal approximators, property given by the Universal Approximation Theorem, which essentially states that any continuous function in any dimension can be represented to arbitrary accuracy by means of an ANN. For this reason, ANNs have a wide range of application, and their use has become ubiquitous in many fields: computer vision, natural language processing, autonomous vehicles, etc. Deep learning algorithms are usually classified according to the amount and type of supervision they get during training and, among all the categories that can be identified, we highlight the supervised and the unsupervised algorithms. They differ in whether they receive the desired solutions in the training set or not.

The aforementioned universal approximation property was exploited in the seminal papers \cite{MEADE19941}, \cite{dissanayake1994neural} and \cite{Lagaris_1998} to introduce a technique to solve partial differential equations (PDEs) by means of ANNs. In recent years there has been a growing interest in approximating the solution of PDEs by means of deep neural networks. They promise to be an alternative to classical methods such as Finite Differences (FD), Finite Volumes (FV) or Finite Elements (FE). For example, the FE technique consists in projecting the solution in some functional space, the Galerkin spaces. Then, by passing to the weak variational formulation and taking the discrete basis, we can find a linear system of equations whose unknowns are the approximated values of the solution as each point of the mesh. In a similar manner, the ANN can be trained to learn data from a physical law that is given by a PDE or a system of PDEs. The idea is quite similar to the classical Galerkin methods, but instead of representing the solution as a projection in some flavour of Galerkin space, the solution is written in terms of ANNs as the composition of nonlinear functions depending on some network weights. As a result, instead of a high dimensional linear system, a high dimensional nonlinear optimization problem is obtained for the ANN weights. This problem must be solved using nonlinear optimization algorithms such as stochastic gradient descent-based methods, e.g., \cite{kingma2014adam}, and/or quasi-Newton methods, e.g., L-BFGS, \cite{doi:10.1137/16M1080173}. More recently, with the advances in automatic differentiation algorithms (AD) and hardware (GPUs), this kind of techniques have gained more momentum in the literature and, currently, the most promising approach is known as physics-informed neural networks (PINNs), see \cite{MENG2020113250} \cite{raissi2017physics}, \cite{doi:10.1137/19M1274067}, \cite{mishra2022estimates}, \cite{doi:10.1063/5.0086649}.

In the last few years, PINNs have shown a remarkable performance. However, there is still some room for improvements within the methodology. One of the disadvantages of PINNs is the lack of theoretical results for the control of the approximation error. Obtaining error estimates or results for the order of approximation in PINNs is a non-trivial task, much more challenging than in classical methods. Even so, the authors in \cite{kolmogorov}, \cite{kdv_bounds}, \cite{mishra2022estimates}, \cite{wPINNs}, \cite{mishra_hiperbolic}, \cite{navier_stokes_mishra} and \cite{MishraBounds} (among others) have derived estimates and bounds for the so-called generalization error considering particular models. Another drawback is the difficulty when imposing the boundary conditions (a fact discussed further later in this section). Nevertheless the use of ANNs has several advantages for solving PDEs: they can be used for nonlinear PDEs without any extra effort; they can be extended to (moderate) high dimensions; and they yield accurate approximations of the partial derivatives of the solution thanks to the AD modules provided by modern deep learning frameworks.

PINNs is not the only approach relying on ANNs to address PDE-based problems. They can be used as a complement for classical numerical methods, for example training the neural network to obtain smoothness indicators, or WENO reconstructions in order for them to be used inside a classical FV method, see \cite{fifth_WENO}, \cite{WENO}. Also ANNs are being used to solve PDE models by means of their backward stochastic differential equation (BSDE) representation as long as the Feynmann-Kàc theorem can be applied, which is the usual situation in computational finance, for example. In \cite{doi:10.1073/pnas.1718942115}, the authors present the so called DeepBSDE numerical methods and their application to the solution of the nonlinear Black-Scholes equation, the Hamilton-Jacobi-Bellman equation, and the Allen-Cahn equation in very high (hundreds of) dimensions. The connection of such method with the recursive multilevel Picard approximations allows the authors to prove that DeepBSDEs are capable of overcoming the so called “curse of dimensionality” for a certain kind of PDEs, see \cite{Picard}, \cite{Jentzen_COD}.

The main goal of the present work is to develop robust and stable deep learning numerical methods for solving nonlinear parabolic PDE models by means of PINNs. The motivation arises from the difficulty of finding and numerically imposing the boundary conditions, which are always delicate and critical tasks both in the classical FD/FV/FE setting and thus also in the ANN setting. The common approach consists in assigning weights to the different terms involved in the loss function, where the selection of this weights must be done heuristically. We introduce a new idea that consists in introducing the loss terms due to the boundary conditions by means of evaluating the PDE operator restricted to the boundaries. In this way the value of such addends is of the same magnitude of the interior losses. Although this is non feasible in the classical PDE solving algorithms, it is very intuitive within the PINNs framework since, by means of AD, we can evaluate this operator in the boundary even in the case it contains normal derivatives to such boundary. Thus, this novel treatment of the boundary conditions in PINNs is the main contribution of this work, allowing to get rid of the heuristic choice of the weights for the contributions of the boundary addends to the loss function that come from the boundary conditions. Further, AD can be naturally exploited to obtain accurate approximations of the partial derivatives of the solution with respect to the input parameters (quantities of much interest in several fields).

Although the proposed methodology could be presented for a wide range of applications, here we will focus on the solution of PDE models for challenging problems appearing the the computational finance field. In particular, we consider the derivative valuation problem in the presence of counterparty credit risk (CCR), which includes in its formulation the so-called x-value adjustments (XVA). This term refers to the different valuation adjustment that arise in the models when the CCR is considered, i.e., when the possibility of default of the parties involved in the transaction is taking into account. These adjustments can come from different sources within a derivative portfolio: credit (CVA), debit (DVA); funding costs (FVA); collateral requirements (ColVA); and capital requirements (KVA), among others. After the $2007$-$2009$ financial crisis, CCR management became of key importance in the financial industry. Several models were developed in order to enrich the classical pricing models by the introduction of risk terms. In this sense, the value adjustments are terms to be added to, or subtracted from, an idealised reference portfolio value, computed in the absence of frictions, in order to obtain the final value of the transaction.

The first works in this topic appeared before the above-mentioned crisis, focusing on analyzing the CVA concept. Some seminal works from this period are \cite{https://doi.org/10.1111/j.1540-6261.1996.tb02712.x}, \cite{Brigo-Masseti-05} and \cite{Cherubini}. After the crisis, the XVA adjustments gained huge attention. The models in which the possibility of default of the parties involved in a transaction were revised by the introduction of the DVA factor, \cite{doi:10.1142/S0219024911006759}, \cite{https://doi.org/10.1111/j.1467-9965.2012.00520.x}. Additionally, the increasingly important role of collateral agreements demands for a portfolio-wide view of valuation by introducing the ColVA factor. In a Black-Scholes economy, \cite{Piterbarg-2010} gives valuation formulas both in the collateralized and uncollateralized case. In addition, generalizations to the case of a multi-currency economy can be found in \cite{Piterbarg-2012}, \cite{Fujii-2010},  \cite{Fujii-2011}, and \cite{doi:10.1137/20M1324375}. Another important aspect for the industry, apart from default risk, is represented by funding costs. Currently, the trading activity is dependent on different sources of liquidity such as the interest rate multi-curve, \cite{multi_curve_ir}, and the old assumption of a unique risk-free interest rate is no longer realistic. In \cite{https://doi.org/10.48550/arxiv.1112.1521}, the FVA is included into a risk-neutral pricing framework for CCR considering realistic settings. Such work is extended in \cite{doi:10.1142/S2345768614500019}, where the effect of Central Clearing Counterparties (CCPs) on funding costs is studied. In this regard, there are many more contributions in obtaining a single risk management framework which includes funding and default risk. In \cite{https://doi.org/10.48550/arxiv.1802.10228} is developed a unified valuation theory that incorporates credit risk, collateralization and funding costs by means of the so-called discounting approach. The authors in \cite{burgard2011partial}, \cite{Burgard-Kjaer-11b} generalize the classical Black-Scholes replication approach to include some of the aforementioned effects. A more general BSDE approach is provided by \cite{https://doi.org/10.1111/mafi.12005}, \cite{Crepey-2020},
\cite{Bichuch_2}, and \cite{Bichuch_1}. In addition, the equivalence between the discounting and BSDE-based approaches is demonstrated in \cite{https://doi.org/10.48550/arxiv.1802.10228}.

Of course, the world of quantitative finance in general, and CCR management in particular, has not been exempt from the advances in deep learning and, nowadays, ANNs are employed for a wide variety of tasks in the industry. Unsupervised ANNs, in both flavours, PINNs and DeepBSDEs, have been recently used for solving challenging financial problems. For example, in \cite{math_Bea} the authors apply PINNs for solving the linear one and two dimensional Black-Scholes equation, and \cite{Weinan-Han-Jentzen-17} introduces the solution of high dimensional Black-Scholes problems using BSDEs. In \cite{https://doi.org/10.48550/arxiv.2005.02633} the authors present a novel computational framework for portfolio-wide risk management problems with a potentially large number of risk factors that makes traditional numerical techniques ineffective. They use a coupled system of BSDEs for XVA which is addressed by a recursive application of an ANN-based BSDE solver. Other relevant works that make use of ANNs for computational finance problems, although not formulated as PDEs, include \cite{horvath2021}, \cite{Savine},  or \cite{liu2021}, for example.

The outline of this paper is as follows. In Section \ref{sec:pinns} we start by revisiting the PINNs framework for solving PDEs. Section \ref{sec:trainingStrategy} introduces the new methodology for the treatment of the boundary conditions in the PINNs setting. In Section \ref{sec:pdes}, the XVA PDE models that we solve in this paper and the adaptation to our PINNs extension are described; more precisely, XVA problems in one and two dimensions, under on Black-Scholes and Heston models. Finally, in Section \ref{sec:experiments}, the numerical experiments that assess the accuracy of the approximation for option prices and their partial derivatives (the so-called Greeks) are presented. 
    
    \section{PINNs} \label{sec:pinns}
    In this section we introduce the so-called PINNs methodology for solving PDEs. The illustration is carried out according to the kind of PDEs that arise in the selected financial problems, i.e., semilinear parabolic PDEs with source terms. Thus, let $\Omega\subset\R^d, d\in\mathbb{N},$ be a bounded, closed and connected domain and $T>0$. Consider the following boundary value problem. Given a function $f\in\mathcal{C}(\mathbb{R})$ and setting $\hat{d}=d+1$, find
$u: (t, x)\in \left[0, T\right]\times \Omega \subset \R^{\hat{d}} \longrightarrow\R$ such that
\begin{align} \label{EQ_parabolic}
    \begin{cases}
        \dfrac{\partial u}{\partial t}(t,x)  + \mathcal{L}[u](t,x) - f(u(t,x)) = 0, & \forall\, (t, x) \in (0, T)\times\openOmega,  \\
        \mathcal{B}[u](t, x) - g(t, x) = 0,&  \forall\, (t, x)\in (0, T)\times\partial\Omega, \\ 
        u(0, x) - u_0(x) = 0,& \forall\, x\in \Omega,
    \end{cases}
\end{align} 
where $\mathcal{L}[\cdot]$ is a strongly elliptic differential operator of second order in the space variables $x$, and $\mathcal{B}[\cdot]$ is a boundary operator defined, for example, by a Dirichlet and/or Neumann boundary conditions. The goal is to approximate this unknown function $u$ by means of a feed-forward neural network, $u_{\theta}(t, x) := u(t, x; \theta),$ where $\theta\in\mathbb{R}^P$ are the network parameters.

\subsection{Feed-forward neural networks} \label{subsec:NN}
    A feed-forward network is a map that transforms an input $y\in\R^{\hat{d}}$ into an output $z\in\R^m$ by means of the composition of a variable number, $L$, of vector-valued functions called layers. These consist of units (neurons), which are the composition of affine-linear maps with scalar non-linear activation functions, \cite{goodfellow}. Thus, assuming a $L$-layer network with $\hat{d}_l$ neurons per layer, it admits the representation
    \begin{equation}
        h(y; \theta) := h_L(\,\cdot\,, \theta^L)\circ h_{L-1}(\,\cdot\,, \theta^{L-1})\circ\cdots\circ h_1(\,\cdot\,, \theta^1)(y),
    \end{equation}
    where, for any $1\leq l \leq L$,
    \begin{equation}
        h_l(z_l; \theta^l) = \sigma_l(W_l z_l + b_l), \quad W_l\in \R^{\hat{d}_{l+1}\times \hat{d}_l}, z_l\in\R^{\hat{d}_l}, b_l\in\R^{\hat{d}_{l+1}},
    \end{equation}
    with $z_1 = y,\,\,\hat{d}_1 = \hat{d}$ and $\hat{d}_L=m$.
    
    Usually (and this is taken as a guideline in this paper) the activation functions are assumed to be the same in all layers except in the last one, where we consider the identity map, $\sigma_L(\,\cdot\,) = Id(\,\cdot\,)$. In addition, taking into account the nature of the problem, it is required that the neural network fulfills the differentiability conditions imposed by \eqref{EQ_parabolic}, requiring sufficiently smooth activation functions such as the sigmoid or the hyperbolic tangent, \cite{shin2020convergence}.
    
    Lastly, it should be noted that a network as the one described above has $\hat{d} + m + \sum_{l=2}^{L-1}\hat{d}_l$ neurons, with parameters $\theta_l = \{W_l, b_l\}$ per layer, yielding a total of
    \begin{equation}
        P = \sum_{l=1}^{L-1}(\hat{d}_l+1)\hat{d}_{l+1}
    \end{equation}
    parameters, which determine the network's capacity.

\subsection{Loss function and training algorithm}
    In order to obtain an approximation of the function $u$ by means of a neural network, $u_{\theta}$, we need to find the network's parameters, $\theta\in\mathbb{R}^P$, that yields the best approximation of \eqref{EQ_parabolic}. This leads to a global optimization problem that can be written in terms of the minimization of a loss function, that measures how good the approximation is. The most popular choice for PINNs' methods is to reduce the problem \eqref{EQ_parabolic} to an unconstrained optimization problem, \cite{dissanayake1994neural}, leading to the family of loss functions involving the $L^2$ error minimization of the interior, initial and boundary residuals. Thus, the loss function, $\mathcal{J(\theta)}$, is defined as
    \begin{equation*} 
        \mathcal{J(\theta)} := \lambda_{\mathcal{I}} \left|\left|\mathcal{R_{\theta}^I}\right|\right|_{L^2((0, T)\times\Omega)}^2 +
        \lambda_{\mathcal{B}} \left|\left| \mathcal{R_{\theta}^B} \right|\right|_{L^2((0, T)\times\partial\Omega)}^2  + \lambda_{\mathcal{O}} \left|\left| \mathcal{R_{\theta}^O} \right|\right|_{L^2(\Omega)}^2,
    \end{equation*}
    or, equivalently,
    \begin{equation} \label{Eq_lossFunction}
        \mathcal{J(\theta)} = \lambda_{\mathcal{I}} \int_{0}^{T}\int_{\Omega}\left|\mathcal{R_{\theta}^I}(t, x)\right|^2\,\text{d}x\text{d}t 
        +\lambda_{\mathcal{B}} \int_{0}^{T} \int_{\partial\Omega}\left| \mathcal{R_{\theta}^B}(t, x) \right|^2\,\text{d}\sigma_x \text{d}t + \lambda_{\mathcal{O}} \int_{\Omega} \left|\mathcal{R_{\theta}^O}(x)\right|^2\,\text{d}x,
    \end{equation}
    where 
    \begin{align}
        \label{interiorResidual}
        \mathcal{R_{\theta}^I}(t, x) &:= \dfrac{\partial u_{\theta}}{\partial t}(t,x) + \mathcal{L}[u_{\theta}](t, x) - f(u_{\theta}(t, x)), & (t, x) \in \,& (0, T)\times \openOmega,\\
        \label{boundaryResidual}
        \mathcal{R_{\theta}^B}(t, x) &:= \mathcal{B}[u_{\theta}](t, x) - g(t, x),&  (t, x)\in \, & (0, T)\times\partial\Omega,\\ 
        \label{initialResidual}
        \mathcal{R_{\theta}^O}(x) &:= u_{\theta}(0, x) - u_0(x),&  x\in\, & \Omega,
    \end{align}
    account for the residuals of the equation, the boundary condition and the initial condition, respectively. The $\lambda_j \in \mathbb{R}^{+},\,\, j\in\{\mathcal{I}, \mathcal{B},  \mathcal{O} \},$ are preset hyperparameters (or updateables during optimization) that allow to impose a weight to each addend of the loss, as can be seen in, e.g., \cite{van2022optimally}, \cite{kgml}. Note that, for the computation of the residuals \eqref{interiorResidual}, \eqref{boundaryResidual}, it is necessary to obtain the derivatives of the neural network with respect to the input space and time variables, well defined under the premise of using sufficiently smooth activation functions. Numerically, they are calculated with the help of AD modules, such those included in Tensorflow, \cite{abadi2016tensorflow}, and Pytorch, \cite{paszke2017zd}. Finally, the strategy followed in PINNs consists of minimizing the loss function \eqref{Eq_lossFunction}, i.e, finding $\theta^{*}\in\Theta$ such that
    \begin{equation} \label{minProblem}
        \theta^{*} = \arg\min_{\theta\in\Theta} \mathcal{J(\theta)},
    \end{equation}
    where $\Theta \subset\mathbb{R}^P$ is the set of admissible parameters.
    
    Except for the simple cases, the integrals appearing in \eqref{Eq_lossFunction} must be computed numerically by means of quadrature rules, \cite{mishra2022estimates}. For this reason, we need to select a set of training points, $\mathcal{P} = \mathcal{P_I} \cup \mathcal{P_B} \cup \mathcal{P_O}$, where
    \begin{align*}
        \mathcal{P_I} &= \{(t_i^{\mathcal{I}}, x_i^{\mathcal{I}})\}_{i=1}^{N_{\mathcal{I}}},\quad (t_i^{\mathcal{I}}, x_i^{\mathcal{I}}) \in (0, T)\times\accentset{\circ}{\Omega}\quad\forall i\in\{1, 2, \cdots, N_{\mathcal{I}}\}, \\
        \mathcal{P_B} &= \{(t_i^{\mathcal{B}}, x_i^{\mathcal{B}})\}_{i=1}^{N_{\mathcal{B}}},\quad (t_i^{\mathcal{B}}, x_i^{\mathcal{B}}) \in (0, T)\times\partial\Omega\quad\forall i\in\{1, 2, \cdots, N_{\mathcal{B}}\}, \\
        \mathcal{P_O} &= \{(0, x_i^{\mathcal{O}})\}_{i=1}^{N_{\mathcal{O}}},\quad\,\,\, x_i^{\mathcal{O}} \in \Omega\quad\forall i\in\{1, 2, \cdots, N_{\mathcal{O}}\}, 
    \end{align*}
    acting as nodes in the quadrature formulas.
    
    Clearly, the choice of the quadrature technique has a direct influence on how these points are selected, and may correspond to, for example, a suitable mesh for a trapezoidal quadrature rule, SOBOL low-discrepancy sequences, a latin hypercube sampling, etc. Moreover, such choice is highly influenced by the problem's time-space dimension, being necessary to use random sampling in high-dimensional domains.
    
    In general terms, we can define the quadrature rule to calculate the integral of a function $\phi: A\subset\R^{\hat{d}} \longrightarrow \R$, as 
    \begin{equation}
        \Phi_M := \sum_{i=1}^M w_i \phi(y_i)
    \end{equation}
    with $\{w_i\}_{i=1}^M\subset\mathbb{R}_{+}$ the weights and $\{y_i\}_{i=1}^M\subset A$ the nodes of the quadrature rule. This allows us to rewrite the loss function \eqref{Eq_lossFunction} taking into account the chosen discretization and quadrature as follows, 
    \begin{equation} \label{Eq_lossFunction_discreticed}
        \mathcal{\hat{J}(\theta)} = \lambda_{\mathcal{I}} \sum_{i=1}^{N_{\mathcal{I}}} w_i^{\mathcal{I}}|\mathcal{R_{\theta}^{\mathcal{I}}}(t_i^I, x_i^I)|^2
        + \lambda_{\mathcal{B}} \sum_{i=1}^{N_{\mathcal{B}}} w_i^{\mathcal{B}}|\mathcal{R_{\theta}^{\mathcal{B}}}(t_i^{\mathcal{B}}, x_i^{\mathcal{B}})|^2
        + \lambda_{\mathcal{O}} \sum_{i=1}^{N_{\mathcal{O}}} w_i^{\mathcal{O}}|\mathcal{R_{\theta}^{\mathcal{O}}}(x_i^{\mathcal{O}})|^2.
    \end{equation}
    
    From now, we will call ``training'' the process of finding the minimum of the problem \eqref{minProblem} with the loss function defined in \eqref{Eq_lossFunction_discreticed}. Even in the case of working with linear PDEs, where the defined functional would be convex, transferring the problem to the parameter space of the neural network yields a high dimensional and highly non-convex problem, \cite{van2022optimally}. As a consequence, the uniqueness of the solution is not guaranteed, and we can only expect to reach a sufficiently low local minima. For this reason, it is common to employ stochastic gradient descent-based methods, such as Adam, \cite{kingma2014adam}, or higher-order quasi-Newton optimizers, such as L-BFGS, \cite{liu1989limited}. In practice, it also implies that a proper choice of model hyperparameters, such as the network size or the learning rate, is essential to achieve a high degree of accuracy. Taking into account what has been explained throughout this section, we detail the steps to find a neural network that approximates the solution of the problem \eqref{EQ_parabolic} in Algorithm \ref{alg:training}.
    
    \begin{algorithm}
        \caption{PINNs' training algorithm}
        \label{alg:training}
        \begin{algorithmic}[1]
            \Require Select a set of training points $\mathcal{P}$, a quadrature rule and an optimization procedure. Define a number of training steps, $N$. Initialize a neural network , $u_{\theta^0}$, with initial parameters $\theta^0$.
            \Ensure Find an approximate local minimum $\theta^*$ of \eqref{minProblem}
            \For{$k=0$, $k{+}{+}$, $k<N$}
                \State $u_{\theta^{k}} \gets u_{\theta^{k}}(t, x)$ \Comment{Evaluate the neural network}
                \State $\mathcal{R}_{\theta^{k}}^{\mathcal{I}}, \mathcal{R}_{\theta^{k}}^{\mathcal{B}}, \mathcal{R}_{\theta^{k}}^{\mathcal{O}} \gets u_{\theta^k}$ \Comment{Compute the residuals}
                \State $\hat{\mathcal{J}}(\theta^k) \gets \mathcal{R}_{\theta^k}^{\mathcal{I}}, \mathcal{R}_{\theta^k}^{\mathcal{B}}, \mathcal{R}_{\theta^k}^{\mathcal{O}}$ \Comment{Compute the loss function}
                \State $\theta^{k+1} \gets \theta^{k}$ \Comment{Apply the optimizer step}
            \EndFor   
        \end{algorithmic}
    \end{algorithm}
    
    Essentially, and except for some particularities, the training process in the case of PINNs is similar to that presented in any other supervised or unsupervised tasks in the field of deep learning. Thus, many of the techniques developed to improve training in such areas can be trivially applied to our case, such as regularization techniques, \cite{goodfellow}, Dropout, \cite{dropout}, transfer learning, \cite{geron}, or other strategies designed to improve the performance of the global optimizer. Once trained, the network serves as an approximate solution to problem \eqref{EQ_parabolic}. It can be evaluated at any point in the domain, and its derivatives can be calculated by AD in few seconds. 
    
    \begin{remark}
        One of the most popular quadrature techniques is Monte Carlo integration. On the one hand, it is a mesh-free method since the points are sampled randomly, making it suitable for high dimensional problems as it does not suffer from the curse of dimensionality. On the other hand, applied to the $L^2$ error expression \eqref{Eq_lossFunction_discreticed}, it gives rise to the mean squared error function, widely used in the deep learning's world.
        
        If we consider a random set of collocation points and define the quadrature weights as
        \begin{equation*}
            w_i^{\mathcal{I}} = \frac{|(0, T)\times\openOmega|}{N_{\mathcal{I}}}, \qquad w_i^{\mathcal{B}} = \frac{|(0, T)\times\partial\Omega|}{N_{\mathcal{B}}},\qquad w_i^{\mathcal{O}} = \frac{|\Omega|}{N_{\mathcal{O}}},
        \end{equation*}
        and taking 
        \begin{equation} \label{LambdasVolume}
            \lambda_{\mathcal{I}} = \frac{\hat{\lambda_{\mathcal{I}}}}{|(0, T)\times\openOmega|}, \qquad \lambda_{\mathcal{B}} = \frac{\hat{\lambda_{\mathcal{B}}}}{|(0, T)\times\partial\Omega|}, \qquad \lambda_{\mathcal{O}} = \frac{\hat{\lambda_{\mathcal{O}}}}{|\Omega|},
        \end{equation}
        with $\hat{\lambda_j}\in \mathbb{R}_{+},\,\,j\in\{\mathcal{I}, \mathcal{B}, \mathcal{O}\}$, then we obtain
        \begin{equation*} \label{Eq_lossFunction_mse}
            \mathcal{\hat{J}(\theta)} = \hat{\lambda_{\mathcal{I}}} MSE_{\mathcal{I}} + \hat{\lambda_{\mathcal{B}}} MSE_{\mathcal{B}} + \hat{\lambda_{\mathcal{O}}} MSE_{\mathcal{O}},
        \end{equation*}
        where
        \begin{gather*}
            MSE_{\mathcal{I}} =  \frac{1}{N_{\mathcal{I}}}\sum_{i=1}^{N_{\mathcal{I}}}|\mathcal{R_{\theta}^{\mathcal{I}}}(t_i^{\mathcal{I}}, x_i^{\mathcal{I}})|^2,\\
            MSE_{\mathcal{B}} =  \frac{1}{N_{\mathcal{B}}}\sum_{i=1}^{N_{\mathcal{B}}} |\mathcal{R_{\theta}^{\mathcal{B}}}(t_i^B, x_i^B)|^2,\\
            MSE_{\mathcal{O}} = \frac{1}{N_{\mathcal{O}}}\sum_{i=1}^{N_{\mathcal{O}}}|\mathcal{R_{\theta}^{\mathcal{O}}}(x_i^{\mathcal{O}})|^2,
        \end{gather*}
        which resembles the loss function employed in most of the works in this topic.
    \end{remark}

    \subsection{Convergence and generalization error bounds}
        With the growth of these methodologies, it is of increasing interest to derive convergence results, as they exist in the finite differences and finite elements world. There are works, such as \cite{shin2020convergence}, in which classical notions of consistency and stability are exploited to prove the strongly convergence of the minimizer to the solution of the linear second-order elliptic or parabolic problem, as the number of collocation points grows. This assumes a random discretization of the domain, together with Monte Carlo integration. 
        
        However, most of the theoretical work on PINNs is dominated by the search for generalization error bounds, where the generalization error, $\mathcal{E}_G(\theta)$, is understood as the total error of the approximated solution, which in our case is given by the square root of the loss function \eqref{Eq_lossFunction}, i.e., 
        \begin{equation*}
            \mathcal{E}(\theta)_G^2 = \mathcal{J(\theta)},
        \end{equation*}
        and depends on the network parameters $\theta\in\Theta$. As discussed in the previous section, evaluating this expression requires the use of numerical integration methods with their respective quadrature points, $\mathcal{P}$. In this sense, the square root of the discretized version of the loss function, given in \eqref{Eq_lossFunction_discreticed}, serves to approximate the generalization error and is also known as training error, $\mathcal{E}_T(\theta, \mathcal{P})$.
        
        Under this setting, we find several papers that attempt to bound the generalization error, for specific problems, in terms of the training error, the chosen quadrature rule, the number of collocation points and the stability of the underlying PDE. For example, such bounds are obtained for the linear Kolmogorov equation, \cite{kolmogorov}, the equation related to the viscous scalar conservation laws and the semi-linear parabolic equation, \cite{mishra2022estimates}, among others. Thus, under existence, uniqueness and regularity assumptions for the semi-linear parabolic case with Lipschitz nonlinearities, the Theorem $3.1$ from \cite{mishra2022estimates} states that the generalization error can be estimated as
        \begin{equation*}
            \mathcal{E}_G \leq C_1\left(\mathcal{E}^{\mathcal{O}}_T + \mathcal{E}^{\mathcal{I}}_T + C_2 (\mathcal{E}^{\mathcal{B}}_T)^{\frac{1}{2}} + (C_q^{\mathcal{O}})^{\frac{1}{2}}N_{\mathcal{O}}^{-\frac{\alpha_{\mathcal{O}}}{2}} + (C_q^{\mathcal{I}})^{\frac{1}{2}}N_{\mathcal{I}}^{-\frac{\alpha_{\mathcal{I}}}{2}} + C_2 (C_q^{\mathcal{B}})^{\frac{1}{4}}N_{\mathcal{B}}^{-\frac{\alpha_{\mathcal{B}}}{4}} \right),
        \end{equation*}
        where $\mathcal{E}^{\mathcal{X}}_T$ are the training errors which verify the relationship $(\mathcal{E}^{\mathcal{X}}_T)^2 = \mathcal{R}_{\theta}^{\mathcal{X}},\,\mathcal{X}\in\{\mathcal{O}, \mathcal{I}, \mathcal{B}\}$. In addition, 
        $C_q^{\mathcal{X}}N_{\mathcal{X}}^{-\alpha_{\mathcal{X}}}$ are the bounds of the quadrature error related to the initial condition, interior domain and boundary, respectively; and $C_1$, $C_2$ are constants that depend on the regularity of the true solution and neural network approximation on the boundary, together with the temporal domain. This result is of special interest because its hypotheses fit within our general problem \eqref{EQ_parabolic} and, furthermore, since we will work with a non-linear contractive source term, the result can be easily applied to the particular problems presented in Section \ref{sec:pdes}.
        
        In a recently published paper, \cite{MishraBounds} present several error bounds in a more abstract framework. Under sufficiently smooth domains, and under the assumptions: $1)$ there exist a neural network that can approximate the solution of the time-dependent PDE at time $T$ with a prescribed tolerance $\epsilon$; and $2)$ the error of the PINNs algorithm can be bounded by means of the error related to its partial derivatives; the following theorem holds. 
        \begin{theorem} (\cite{MishraBounds})
            Let $r$, $s\,\in \mathbb{N}$, let $u\,\in\,\mathcal{C}^{(s, r)}([0, T]\times \Omega)$ be the solution of the abstract time-dependent PDE with initial condition $u_0 \in L^2(\Omega)$ and let the above assumptions be satisfied. There exists a constant $C(s, r) > 0$ such that for every $M \in \mathbb{N}$ and $\epsilon > 0$ there exist a neural network $u_{\theta}: [0, T]\times \Omega \longrightarrow \R$, with the hyperbolic tangent as activation function, for which it holds that,
            \begin{equation*}
                ||u_{\theta} - u||_{L^{q}([0, T]\times \Omega)} \leq C(||u||_{C^0}M^{-s} + \epsilon),
            \end{equation*}
            where $M$ is the number of spatial intervals chosen in the discretization.
        \end{theorem}
        Moreover, this theorem includes an additional result in which the $L^2$-norm of the operator applied to the neural network is bounded, and both statements together imply that there exists a neural network for which the generalization error and the PINN's loss function can be made as small as possible. Since our framework is embedded within this abstract formulation, such result ensures a solid theoretical foundation for our work.

    \section{Novel treatment of boundary conditions} \label{sec:trainingStrategy}
    Ideally, the loss function correctly captures how far away we are from the exact solution of the problem and how well the boundary restrictions are fulfilled, so that the optimization algorithm can get us close to a good local minima, at least. However, in practice, this situation is not always reproduced when applying numerical methods. In the case of PINNs we also have this problem and, although the reasons why this happens are poorly understood, previous works, such as \cite{kgml}, point to the fact that training is focused on getting a small PDE residual in the interior domain, while leading to large errors in the fitting of the boundary conditions. This suggest that the contribution of the some boundary errors vanishes.

In most of the works on this topic, this problem is usually solved by introducing the lambda weights seen before, which preponderate the contribution of each of the terms involved in the elaboration of the loss function \eqref{Eq_lossFunction_discreticed}. The optimal choice of this weights is of paramount importance for the algorithm. The main drawback of this methodology is that the choice of these values is problem-dependent and in most situations is carried out heuristically, \cite{kgml}.

We identify that the introduction of the overriding factors is mainly driven by two features. On the one hand, we encounter the problem that the integrals involved in the loss function present different domain dimensionality, i.e., introduce different magnitudes of volume. The integral referring to the residual in the interior of the domain involves a $\hat{d}$-volume, while the integrals associated with the initial and boundary residuals involve a ($\hat{d}$-$1$)-volume. 

An easy solution to solve this situation is to force these lambdas to be inversely proportional to the volume of the each integral's domain considered (as we have shown for the Monte Carlo case). Then, taking into account \eqref{LambdasVolume}, we rewrite the discrete loss function \eqref{Eq_lossFunction_discreticed} as

\begin{equation} \label{DiscretizedLossFunctionVols}
    \mathcal{\hat{J}(\theta)} = \frac{\hat{\lambda_{\mathcal{I}}}}{|(0, T)\times\openOmega|} \sum_{i=1}^{N_{\mathcal{I}}} w_i^{\mathcal{I}}|\mathcal{R_{\theta}^{\mathcal{I}}}(t_i^I, x_i^I)|^2
    + \frac{\hat{\lambda_{\mathcal{B}}}}{|(0, T)\times\partial\Omega|}\sum_{i=1}^{N_{\mathcal{B}}} w_i^{\mathcal{B}}|\mathcal{R_{\theta}^{\mathcal{B}}}(t_i^{\mathcal{B}}, x_i^{\mathcal{B}})|^2 
    + \frac{\hat{\lambda_{\mathcal{O}}}}{|\Omega|} \sum_{i=1}^{N_{\mathcal{O}}} w_i^{\mathcal{O}}|\mathcal{R_{\theta}^{\mathcal{O}}}(x_i^{\mathcal{O}})|^2.
\end{equation}

On the other hand, the magnitude of the contributions to the loss function can differ in several orders, i.e., there are addends which are negligible with respect to others, leading to a worse local minima in the training, or the need to extend training time. In general, there are two possible situations that can occur simultaneously in a boundary value problem. One of them is that we can find residuals with large relative losses as the beginning of training. As a consequence, they can cause longer training times, as in the early stages of training the loss function only provides information regarding such losses. The other possibility is that we can find boundaries in which the residuals exhibit relatively much smaller values, so their contribution to the loss function is, in many cases, negligible. As a consequence, such constrains could not provide information to the training.

In order to avoid the arbitrary selection of the loss function weights, it is essential to reduce the differences in magnitude among residuals. For this reason, we propose, for the first time to the best of our knowledge, a novel approach which overcomes these weights' issue. It is based on reformulating, whenever possible, the residuals related to Dirichlet or Neumann (Robin, higher order derivatives) conditions. This reformulation relies on taking as a residual not the boundary condition itself but the resulting PDE restricted to the corresponding boundary. This will produce losses of an order of magnitude similar to that produced by the interior residual, once these quantities are dimensionless.

Thus, most of the Neumann, Robin or higher order derivative boundary residuals we will work with can be written in this form. It suffices to substitute the condition into the PDE of the interior domain and impose the resulting equation on the related boundary residual. However, it will only be possible to impose Dirichlet conditions in this way when they naturally occur at the boundary, i.e., when the Dirichlet condition arises from solving the differential equation that results at the boundary. 

As an illustrative example, let us consider a particular case of the parabolic problem defined in \eqref{EQ_parabolic}, where Dirichlet and Neumann boundary conditions are presented. Under the spatial domain $\Omega = \prod_{i=1}^d\left[x_i^{min}, x_i^{max}\right]$, the upper boundaries
\begin{equation*}
    \Gamma_{x_i}^+ = (x_1^{min}, x_1^{max})\times\dots\times \{x_i^{max}\}\times\cdots\times (x_d^{min}, x_d^{max}),\quad i=1,\cdots d,
\end{equation*}
and the lower boundaries
\begin{equation*}
    \Gamma_{x_i}^0 = (x_1^{min}, x_1^{max})\times\dots\times \{x_i^{min}\}\times\cdots\times (x_d^{min}, x_d^{max}),\quad i=1,\cdots d,
\end{equation*}
we want to find the parameters of an ANN $u_{\theta}$ in order to make it verify 
\begin{align} \label{examplePDE}
    \begin{cases}
        \dfrac{\partial u_{\theta}}{\partial t} + \displaystyle\sum_{i, j=1}^d a_{ij}\dfrac{\partial^2 u_{\theta}}{\partial x_i\partial x_j} + \displaystyle\sum_{i=1}^d b_i \dfrac{\partial u_{\theta}}{\partial x_i} + f(u_{\theta}) = 0, & \text{in}\,\, (0, T)\times\openOmega, \\
        \dfrac{\partial u_{\theta}}{\partial x_i} - g_i=0 & \text{in}\,\,\Gamma_i^+=(0, T)\times\Gamma_{x_i}^+,\,\,\,i=1,\dots d,\\
        u_{\theta} - h_i = 0 & \text{in}\,\, \Gamma_i^0 = (0, T)\times \Gamma_{x_i}^0,\,\,\, i=1,\dots d,\\
        u_{\theta} - u_0 = 0 & \text{in}\,\,\Omega,
    \end{cases}
\end{align}
where $\{a_{ij}\}_{i, j=1}^d \subset \R,\,\, \{b_i\}_{i=1}^d\subset\R\setminus\{0\}$, and $g_i\in\mathcal{C}(\Gamma_i^+, \R) $, $h_i \in \mathcal{C}(\Gamma_i^0, \R),\,\,\,i=1,\dots d$. For example, when defining the residuals associated with the Neumann conditions, the usual approach is to take the condition itself as the residual, i.e.
\begin{equation*}
    \mathcal{R_{\theta}}^{\Gamma_i^+} = \dfrac{\partial u_{\theta}}{\partial x_i} - g_i,\quad i=1,\dots d.
\end{equation*}
Alternatively, in our proposal, we plug the Neumann condition into the PDE and impose the resulting equation as a residual, obtaining
\begin{equation*}
    \mathcal{R_{\theta}}^{\Gamma_i^+} = \dfrac{\partial u_{\theta}}{\partial t} + \displaystyle\sum_{i, j=1}^d a_{ij}\dfrac{\partial^2 u_{\theta}}{\partial x_i\partial x_j} + \displaystyle\sum_{\substack{j=1 \\j\neq i}}^d b_j \dfrac{\partial u_{\theta}}{\partial x_j} + b_i g_i + f(u_{\theta}),\quad i=1,\cdots d.
\end{equation*}
For Dirichlet conditions, the proposed strategy can only be applied when $h_i$ verifies the PDE and the initial condition of \eqref{examplePDE} at the boundary $\Gamma_i^0$. In such cases we can define the residual in the same way as the residual of the PDE, i.e.,
\begin{equation*}
    \mathcal{R_{\theta}}^{\Gamma_i^0} = \dfrac{\partial u_{\theta}}{\partial t} + \displaystyle\sum_{i, j=1}^d a_{ij}\dfrac{\partial^2 u_{\theta}}{\partial x_i\partial x_j} + \displaystyle\sum_{i=1}^d b_i \dfrac{\partial u_{\theta}}{\partial x_i} + f(u_{\theta}),\quad i=1,\dots d.
\end{equation*}
Because of that, such kind of Dirichlet conditions does not even need to be included as boundary residuals. Depending on the quadrature scheme employed, it would be enough to force the existence of interior domain collocation points on such boundary.

\begin{remark}
    As a summary, we have first briefly described the main problems that lead to the introduction of additional weights in the loss function. Then we have proposed a new treatment of the boundary residuals that allows to avoid such weights. When we deal with derivative-based boundary conditions, the related residuals are defined by taking the equation resulting from substituting the boundary conditions in the PDE. For residuals associated with Dirichlet boundary conditions, we can impose the PDE itself as a boundary residual as long as it arises naturally on such boundary.
\end{remark}

    \section{Application to problems in computational finance} \label{sec:pdes}
    In this section we present the PDE formulation of the particular problems we will address in this work. We focus on some relevant (and challenging) state-of-the-art problems appearing in computational finance, specifically, in the area of the CCR assessment. Thus, we consider the valuation of some financial derivatives when accounting for such a risk, namely the pricing of different \emph{risky} European option under the Black-Scholes and Heston model. All of them are extensions of the risk-free derivative pricing models to a formulation that takes into account the effects of bilateral default risk and the funding costs, i.e., which includes CVA, DVA and FVA adjustments, following the approach of \cite{burgard2011partial}. We chose this methodology for its simplicity, but any more complex extension, such as \cite{BRIGO2019}, can fit into our framework.

\subsection{General pricing problem formulation}
    Let us consider a derivative contract $\hat{V}$ on $d\geq 1$ spot assets, $S\in \R_{+}^d$, between two parties, the seller B and its counterparty C, where both may default. We assume that the default of either B or C does not affect $S$. Such derivative pays the seller B the amount $H(S)\in\R$ at maturity $T$. In addition, let $V$ the same derivative between two parties that cannot default, i.e., the non-risky derivative value.
    
    Under the described setting, if either the seller or the counterparty defaults, the International Swaps and Derivative Association (ISDA) Master Agreement determines that the value of the derivative is fixed by a Mark-to-Market rule $M$, which is chosen to be either $\hat{V}$ or $V$, adjusted by means of $R_B, R_C \in [0,1]$, the recovery rates on $M$ if seller or counterparty defaults, respectively. Considering $r$ as the risk-free interest rate, $r_B$ the seller's bond yield and $r_C$ the counterparty's bond yield. Following \cite{burgard2011partial} and \cite{salvador2021total} we can define the B and C's default intensities, $\lambda_B$ and $\lambda_C$, by means of the spread between their bond yields and the risk-free interest rate, i.e., 
    \begin{equation*}
         \lambda_B = r_B - r,\quad \lambda_C = r_C - r.
    \end{equation*}
    In addition, the seller's funding rate for borrowed cash $r_F$ is considered. If the derivative can be used as collateral, $r_F=r$ is taken, while if collateral cannot be used as collateral, $r_F = r + (1 - R_B)\lambda_B$ is taken. In this regard, we define the funding spread $s_F$ as
    \begin{equation*}
        s_F = r_F - r.
    \end{equation*}
    
    From now on, we establish the Mark-to-Market rule $M=\hat{V}$ and that the derivative cannot be used as collateral, so a non-linear PDE model for $\hat{V}$ is obtained. It follows the general definition
    \begin{equation} \label{GeneralRiskyPDE}
        \begin{cases}
            \dfrac{\partial \hat{V}}{\partial t} + \mathcal{L}[\hat{V}] + f(\hat{V}) = 0, \\ 
            \hat{V}(0, S) - H(S) = 0,
        \end{cases}
    \end{equation}
    where $t$ is the time to maturity variable, $\mathcal{L}$ the differential elliptic operator defined by the chosen problem, and $f$ the non-linear source term given by
    \begin{equation}
        f(\hat{V}) = \lambda_B (1 - R_B)\min \Bigl\{\hat{V}, 0\,\Bigr\} + \lambda_C (1 - R_C) \max\Bigl\{\hat{V}, 0\,\Bigr\} + s_F \max\Bigl\{\hat{V}, 0\,\Bigr\}.
    \end{equation}
    In addition, the derivative value without considering counterparty risk, $V$, obeys the PDE
    \begin{equation} \label{GeneralRiskfreePDE}
        \begin{cases}
            \dfrac{\partial V}{\partial t} + \mathcal{L}[V] = 0, \\ 
            V(0, S) - H(S) = 0.
        \end{cases}
    \end{equation}

\subsection{Specific pricing problem formulation}
    Having defined the general context of the financial problems to be addressed, we are in a position to present the boundary value problems obtained in each specific case, as well as their adaptation to the methodology presented in Section \ref{sec:pinns}.

    \subsubsection{European option under the Black-Scholes model} \label{subsec:blackScholes}
        We consider an European option with strike $K\in\R$ and maturity $T>0$. Let $S$ the underlying stock value, $\sigma$ the volatility in $S$ and $r_R$
        the stock repo rate minus the dividend yield under the Black-Scholes model. The option price, $\hat{V}$, is given by equation \eqref{GeneralRiskyPDE} taking the elliptic operator as
        \begin{equation}
            \label{BSoperator}
            \mathcal{L} = -\frac{\sigma^2 S^2}{2}\dfrac{\partial^2}{\partial S^2}-r_RS\dfrac{\partial}{\partial S} + r \mathcal{I},
        \end{equation}
        and the initial condition the vanilla payoff,
        \begin{equation}    
            \label{vanilla_payoff}
            H(S) = \max \Bigl\{\alpha \left(S - K\right), 0 \,\Bigr\},
        \end{equation}
        with $(t, S) \in [0, T]\times [0, +\infty)$ and $\alpha\in\{-1, 1\}$ for put and call options, respectively. In this case, an analytic solution for \eqref{BSoperator}-\eqref{vanilla_payoff} is known, \cite{burgard2011partial},
        \begin{equation} \label{eq:riskyBSFormula}
            \widehat{\mathcal{BS}}(t, S) = \mathcal{BS}(t, S)\exp{\Bigl\{-(\lambda_B(1 - r_B) + \lambda_C(1-r_C))\,t\Bigr\}}, 
        \end{equation}
        where $\mathcal{BS}(\cdot, \cdot)$ is the solution of the classical Black-Scholes equation:
        \begin{equation} \label{eq:BSFormula}
            \mathcal{BS}(t, S) = \alpha S \exp{\Bigl\{{-(r - r_R)t}\Bigr\}} \Phi(\alpha \zeta_1) - \alpha K \exp{\Bigl\{-rt\Bigr\}} \Phi(\alpha \zeta_2),
        \end{equation}
        with $\Phi(\cdot)$ the cumulative distribution function of a standard normal variable, and
        \begin{equation*}
            \zeta_1 = \dfrac{\log(S / K) + (r_R + 0.5\,\sigma^2) t}{\sigma\sqrt{\tau}},\qquad \zeta_2 = \zeta_1 - \sigma \sqrt{t}.
        \end{equation*}
        
        In order to apply the methodology introduced in Section \ref{sec:pinns} for its resolution, it is necessary to carry out a truncation of the semi-infinite domain $[0, +\infty)$ into $[0, S_{max}]$. This step enforces us to include boundary conditions when  $S=S_{max}$. For the left boundary, $\Gamma^0 = (0, T)\times\{0\}$ , it is sufficient to substitute $S=0$ in the equation \eqref{GeneralRiskyPDE} with \eqref{BSoperator}-\eqref{vanilla_payoff}, obtaining
        \begin{equation} \label{eq:BSLeftBoundary}
            \dfrac{\partial \hat{V}}{\partial t} + r\hat{V} + f(\hat{V}) = 0,
        \end{equation}
        which can be imposed as the following Dirichlet boundary condition,
        \begin{equation}\label{eq:BSLeftDirichlet}
            \hat{V}(t, 0) = \dfrac{|\alpha - 1|}{2}K \exp{\Bigl\{-(r + \lambda_B(1 - r_B) + \lambda_C(1 - r_C))\, t\,\Bigr\}}.
        \end{equation}
        Taking into account that
        \begin{equation} \label{eq:BSRightBoundary}
            \lim_{S\to\infty} \dfrac{\partial^2 \hat{V}}{\partial S^2} = 0,
        \end{equation}
        we can consider such linear boundary condition for the right boundary, $\Gamma^+ = (0, T)\times\{S_{max}\}$, when $S_{max}$ is large enough. Examples of application can be viewed in, e.g. \cite{chen2019penalty}. 
        
        Thus, for $\Omega = [0, S_{max}]$, the European option considering CCR from above verifies the following boundary value problem. Find $\hat{V}: [0, T]\times\Omega\subset \R^2 \longrightarrow \R$ such that
        \begin{equation} \label{eq:RiskyBS}
            \begin{cases}
                \dfrac{\partial\hat{V}}{\partial t} - \dfrac{\sigma^2 S^2}{2}\dfrac{\partial^2\hat{V}}{\partial S^2}-r_RS\dfrac{\partial\hat{V}}{\partial S} + r \hat{V} + f(\hat{V}) = 0, & \text{in}\,\,(0, T)\times\openOmega, \\
                \hat{V} - \dfrac{|\alpha - 1|}{2}K \exp{\Bigl\{-(r + \lambda_B(1 - r_B) + \lambda_C(1 - r_C))\, t\,\Bigr\}} = 0,& \text{in}\,\, \Gamma^0, \\
                \dfrac{\partial^2 \hat{V}}{\partial S^2} = 0,& \text{in}\,\, \Gamma^+,\\
                \hat{V} - \max \Bigl\{\alpha \left(S - K\right), 0 \,\Bigr\} = 0, & \text{in}\,\, \{0\}\times\Omega.
            \end{cases}
        \end{equation}
         Moreover, it is straightforward to prove that the European option without considering counterparty risk verifies the equation \eqref{eq:RiskyBS} by taking $\lambda_B = \lambda_C = 0$ or, equivalently, taking $f=0$.
        
         Such formulations fits into problem \eqref{EQ_parabolic} so we can apply everything explained in Section \ref{sec:pinns} to solve it. In order to do this, we consider a discretization of the domain
         \begin{equation*}
            [0, T]\times\Omega = \left((0, T)\times \openOmega\right)\, \cup\,\Gamma^0\,\cup\,\Gamma^+ \cup\,\Bigl(\{0\}\times\Omega\Bigr),  
         \end{equation*}
         by a uniform discretization of each of the presented subsets. For the sake of clarity, we present here a uniform mesh throughout the domain. Thus, calling $N_S$ the number of steps 
        in the $S$-direction and $N_T$ the number of time steps, we take the grids,
        \begin{align*}
             \mathcal{P}_{\mathcal{I}}&=\bigl\{(t_i, S_j)\bigr\}_{i, j = 1}^{N_T, N_S-1} = \bigl\{\left(i\,\Delta_T, j\,\Delta_S\right) : i=1,\dots N_T,\,\, j=1,\dots N_S-1\bigr\}, \\
             \mathcal{P}_{\Gamma^0} &= \bigl\{(t_i, 0)\bigr\}_{i=1}^{N_T}= \bigl\{\left(i\,\Delta_T, 0\right):  i=1,\dots N_T\bigr\}, \\
            \mathcal{P}_{\Gamma^+} &=\bigl\{(t_i, S_{max})\bigr\}_{i=1}^{N_T} = \bigl\{\left(i\,\Delta_T, S_{max}\right): i=1,\dots N_T\bigr\}, \\
            \mathcal{P}_{\mathcal{O}}&=\bigl\{(0, S_j)\bigr\}_{j = 0}^{N_S} = \bigl\{\left(0, j\,\Delta_S\right) :  j=,0 \dots N_S\bigr\},
        \end{align*}
        with $\Delta_S = S_{max} / N_S$ and $\Delta_T = T / N_T$ the steps size. Therefore, the set of collocation points of the problem is given by
        \begin{equation}
          \mathcal{P} = \mathcal{P_I}\cup\mathcal{P}_{\Gamma^0}\cup\mathcal{P}_{\Gamma^+}\cup\mathcal{P_{\mathcal{O}}},  
        \end{equation}
        with size 
        \begin{equation}
          N_{\mathcal{P}} = |\mathcal{P}| = (N_T + 1) \times (N_S + 1).
        \end{equation}
        
        To approximate the desired solution we consider a neural network, $\hat{V_{\theta}}: [0, T]\times\Omega\subset\R^2 \longrightarrow \R$, with $l$ hidden layers. Without loss of generality, we assume that the number of neurons per hidden layer is the same, $\beta$. Based on the boundary value problem given in \eqref{eq:RiskyBS}, we choose the network residuals taking into account our proposal to solve the aforementioned training issues. Thus, since on the boundary $\Gamma^0$ the Dirichlet condition arises naturally, we can use the expression \eqref{eq:BSLeftBoundary} as boundary residual. Moreover, on the boundary $\Gamma^+$ we have a higher-order derivative condition, so we can substitute this condition, \eqref{eq:BSRightBoundary}, into the equation \eqref{GeneralRiskyPDE} in order to impose such residual in the same way as we explain in Section \ref{sec:trainingStrategy}. Applying these considerations, we obtain the following residuals,
        \begin{align}\label{BSResidualInterior}
            \mathcal{R}_{\theta}^{\mathcal{I}} &= \dfrac{\partial\hat{V_{\theta}}}{\partial t} - \dfrac{\sigma^2 S^2}{2}\dfrac{\partial^2\hat{V_{\theta}}}{\partial S^2}-r_RS\dfrac{\partial\hat{V_{\theta}}}{\partial S} + r \hat{V_{\theta}} + f(\hat{V_{\theta}}),& \text{in}\,\, & (0, T)\times\openOmega,
            \\ \label{BSResidualBoundary0}
            \mathcal{R}_{\theta}^{\Gamma^0} &= \dfrac{\partial \hat{V_{\theta}}}{\partial t} + r\hat{V_{\theta}} + f(\hat{V_{\theta}}),&\text{in}\,\,&\Gamma^0, \\  \label{BSResidualBoundary+}
            \mathcal{R}_{\theta}^{\Gamma^+} &= \dfrac{\partial\hat{V_{\theta}}}{\partial t} - r_RS\dfrac{\partial \hat{V_{\theta}}}{\partial S} + r\hat{V_{\theta}} + f(\hat{V_{\theta}}),& \text{in}\,\, &\Gamma^+, \\ 
            \mathcal{R}_{\theta}^{\mathcal{O}} &= \hat{V_{\theta}} - \max \Bigl\{\alpha \left(S - K\right), 0\, \Bigr\}, & \text{in}\,\, & \{0\}\times\Omega.
        \end{align}
        Using these residuals, the loss function is defined in the same way as in \eqref{DiscretizedLossFunctionVols} by taking the lambda weights equal to one and the quadrature weights corresponding to the trapezoidal rule,
        \begin{equation} \label{eq:lossBS1d}
            \begin{split}
                \mathcal{\hat{J}(\theta)} =& \frac{\Delta_T\Delta_S}{4|(0, T)\times\openOmega|}\left(\mathcal{R_{\theta}^{\mathcal{I}}}(t_1, S_1)^2 + \mathcal{R_{\theta}^{\mathcal{I}}}(t_1, S_{N_S-1})^2 + \mathcal{R_{\theta}^{\mathcal{I}}}(t_{N_T}, S_1)^2 + \mathcal{R_{\theta}^{\mathcal{I}}}(t_{N_T}, S_{N_S-1})^2\right)  \\
                +& \frac{\Delta_T\Delta_S}{2|(0, T)\times\openOmega|}\left[
                \sum_{i=1}^{N_T-1}\left( 
                \mathcal{R_{\theta}}^{\mathcal{I}}(t_i, S_1)^2 + \mathcal{R_{\theta}}^{\mathcal{I}}(t_i, S_{N_S-1})^2 
                \right)+ 
                \sum_{j=1}^{N_S-2} \left(\mathcal{R_{\theta}}^{\mathcal{I}}(t_1, S_j)^2 +  \mathcal{R_{\theta}}^{\mathcal{I}}(t_{N_T}, S_j)^2 
                \right)
                \right] \\
                +& \frac{\Delta_T\Delta_S}{|(0, T)\times\openOmega|} \sum_{i=1}^{N_T-1}\sum_{j=1}^{N_S-2}  \mathcal{R_{\theta}^{\mathcal{I}}}(t_i, S_j)^2 + 
                \frac{\Delta_T}{2|\Gamma^0|}\left(
                \mathcal{R_{\theta}}^{\Gamma^0}(t_1)^2 + \mathcal{R_{\theta}}^{\Gamma^0}(t_{N_T})^2
                + 2\sum_{i=1}^{N_T-1} \mathcal{R_{\theta}}^{\Gamma^0}(t_i)^2
                \right) \\
                +&\frac{\Delta_T}{2|\Gamma^+|}\left(
                \mathcal{R_{\theta}}^{\Gamma^+}(t_1)^2 + \mathcal{R_{\theta}}^{\Gamma^+}(t_{N_T})^2
                + 2\sum_{i=1}^{N_T-1} \mathcal{R_{\theta}}^{\Gamma^+}(t_i)^2
                \right) 
                + \frac{\Delta_S}{2|\Omega|}\left(
                \mathcal{R_{\theta}}^{\mathcal{O}}(S_0)^2 + \mathcal{R_{\theta}}^{\mathcal{O}}(S_{N_S})^2
                \right) \\
                +&\frac{\Delta_S}{|\Omega|} 
                \sum_{j=1}^{N_S-1} \mathcal{R_{\theta}}^{\mathcal{O}}(S_j)^2.
            \end{split}
        \end{equation}
    
    \subsubsection{European basket option under the Black-Scholes model} \label{subsec:basket}
        Next, we present a European basket option driven by two assets, $S_1$ and $S_2$, with strike $K\in\R$ and maturity $T>0$. As we did before, for each asset we consider its volatility $\sigma_i$, and its repo rate minus dividend yield $r_{R_{i}} = q_i - \gamma_i $, with $i=1$ for the first asset and $i=2$ for the second. Further, we define the correlation between assets, $\rho$, verifying that $|\rho|\leq 1.$ The basket options' price, $\hat{V}$, is given by the equation \eqref{GeneralRiskyPDE} taking the elliptic operator
        \begin{equation} \label{basketOperator}
            \mathcal{L} = -\dfrac{\sigma_1^2 S_1^2}{2}\dfrac{\partial^2}{\partial S_1^2} - 
            \rho\sigma_1\sigma_2 S_1 S_2\dfrac{\partial^2}{\partial S_1 \partial S_2}
            - \dfrac{\sigma_2^2 S_2^2}{2}\dfrac{\partial^2}{\partial S_2^2} - r_{R_1}S_1\dfrac{\partial}{\partial S_1} - r_{R_2} S_2 \dfrac{\partial}{\partial S_2} + r \mathcal{I},
        \end{equation}
        with the initial condition the derivative's payoff . In our case, we work with two challenging and practically appearing payoffs, namely, the arithmetic average payoff,
        \begin{equation} \label{eq:average_payoff}
            H(S_1, S_2) = \max\biggl\{\alpha\left(\dfrac{S_1 + S_2}{2} - K\right), 0\,\biggr\},
        \end{equation}
        and the \textit{worst-of} payoff:
        \begin{equation} \label{eq:worst_of_payoff}
            H(S_1, S_2) = \max\Bigl\{\alpha\left( \min\Bigl\{S_1, S_2\Bigr\} - K \right), 0\,\Bigr\}.
        \end{equation}
        It is worth noting that in the case of the worst-of risk-free option, an analytical solution is known, see for example \cite{STULZ}, \cite{johnson_1987}.
        
        The spatial domain is a Cartesian product of semi-infinite intervals, $[0, +\infty)\times [0, +\infty)$, and, for its numerical resolution, each interval is truncated, obtaining $\Omega = [0, S_{1, max}]\times [0, S_{2, max}]$. Moreover, additional conditions must be imposed on the boundaries $\Gamma_1^0 = (0, T)\times\{0\}\times[0, S_{2, max}]$, $\Gamma_2^0 = (0, T)\times (0, S_{1, max}]\times\{0\}$, $\Gamma_1^+ = (0, T)\times\{S_{1, max}\}\times(0, S_{2, max})$ and $\Gamma_2^+ = (0, T)\times (0, S_{1, max}]\times\{S_{2, max}\}$. For the lower boundaries it is possible to impose Dirichlet conditions which again arise naturally. At each boundary, $\Gamma_i^0,\,\,i\in \{1, 2\}$, we substitute $S_i = 0$ obtaining
        \begin{equation} \label{eq:boundary_basket_0}
            \dfrac{\partial\hat{V}}{\partial t} - \dfrac{\sigma_j^2 S_j^2}{2}\dfrac{\partial^2\hat{V}}{\partial S_j^2} - r_{R_j} S_j \dfrac{\partial \hat{V}}{\partial S_j} +r\hat{V} + f(\hat{V}) = 0,\quad j\neq i,
        \end{equation}
        which, with the initial condition \eqref{eq:average_payoff}, gives rise to the one-dimensional risky Black-Scholes equation \eqref{eq:riskyBSFormula} depending on the other underlying. If we consider the initial condition \eqref{eq:worst_of_payoff} we can impose the expression \eqref{eq:BSLeftDirichlet}, since the initial condition does not depend on the remaining underlying.
        
        For the upper boundaries $\Gamma_i^+,\,\,i\in \{1, 2\}$, we work with the linear boundary condition \eqref{eq:BSRightBoundary}, used for the non-risky cases in, e.g.,  \cite{Randall2000PricingFI}, since the qualitative behaviour of the solution does not change in the limit.
    
        Thus, the European arithmetic average basket option considering CCR verifies the boundary value problem of finding $\hat{V}: [0, T]\times\Omega\subset\R^3\longrightarrow\mathbb{R}$ such that
        \begin{equation} \label{eq:RiskyAverageBasket}
            \begin{cases}
                \!\begin{aligned}
                    \dfrac{\partial\hat{V}}{\partial t}  - \dfrac{\sigma_1^2 S_1^2}{2}\dfrac{\partial^2\hat{V}}{\partial S_1^2} - & 
                    \rho\sigma_1\sigma_2 S_1 S_2\dfrac{\partial^2\hat{V}}{\partial S_1 \partial S_2}
                    - \dfrac{\sigma_2^2 S_2^2}{2}\dfrac{\partial^2\hat{V}}{\partial S_2^2} \\ 
                    - & r_{R_1}S_1\dfrac{\partial\hat{V}}{\partial S_1} 
                    - r_{R_2} S_2 \dfrac{\partial\hat{V}}{\partial S_2} + r\hat{V} + f(\hat{V}) = 0,
                \end{aligned} & \text{in}\,\,(0, T)\times\openOmega, \\
                \hat{V} - \widehat{\mathcal{BS}}_i = 0,& \text{in} \,\, \Gamma_i^0,\,\,i=1,2, \\
                \dfrac{\partial^2\hat{V}}{\partial S_i^2} = 0, &  \text{in}\,\, \Gamma_i^+,\,\,i=1,2, \\
                \hat{V} -  \max\biggl\{\alpha\left(\dfrac{S_1 + S_2}{2} - K\right), 0\,\biggr\}= 0, &  \text{in}\,\, \{0\}\times\Omega,
            \end{cases}
        \end{equation}
        where $\widehat{\mathcal{BS}}_i$ refers to the Black-Scholes formula \eqref{eq:riskyBSFormula} applied to $S_i, \,\,i\in\{1, 2\}$.
        
        The European worst-of basket option with counterparty risk verifies the value problem of finding the function $\hat{V}: [0, T]\times\Omega\subset\R^3\longrightarrow\mathbb{R}$ such that
        \begin{equation} \label{eq:RiskyWorstOf}
            \begin{cases}
                \!\begin{aligned}
                    \dfrac{\partial\hat{V}}{\partial t}  - \dfrac{\sigma_1^2 S_1^2}{2}\dfrac{\partial^2\hat{V}}{\partial S_1^2} - & 
                    \rho\sigma_1\sigma_2 S_1 S_2\dfrac{\partial^2\hat{V}}{\partial S_1 \partial S_2}
                    - \dfrac{\sigma_2^2 S_2^2}{2}\dfrac{\partial^2\hat{V}}{\partial S_2^2} \\ 
                    - & r_{R_1}S_1\dfrac{\partial\hat{V}}{\partial S_1} 
                    - r_{R_2} S_2 \dfrac{\partial\hat{V}}{\partial S_2} + r\hat{V} + f(\hat{V}) = 0,
                \end{aligned} & \text{in}\,\,(0, T)\times\openOmega, \\
                \hat{V} - \dfrac{|\alpha - 1|}{2}K \exp\{-(r + \lambda_B(1 - r_B) + \lambda_C(1 - r_C)) t\} = 0,& \text{in}\,\, \Gamma_i^0,\,\,i=1,2, \\
                \dfrac{\partial^2\hat{V}}{\partial S_i^2} = 0, &  \text{in} \,\, \Gamma_i^+,\,\,i=1,2,  \\
                \hat{V} - \max\Bigl\{\alpha\left( \min\Bigl\{S_1, S_2\Bigr\} - K \right), 0\,\Bigr\}= 0, & \text{in}\,\,\{0\}\times \Omega.
            \end{cases}
        \end{equation}
        Similarly to the one-dimensional problem seen before, the same boundary problems are valid for the associated risk-free options by taking $\lambda_B = \lambda_C = 0$.
        
        Both formulations fit into the problem \eqref{EQ_parabolic}, and can be applied as discussed in the Section \ref{sec:pinns}. Thus, we consider a discretization of the domain
        \begin{equation*}
            [0, T]\times\Omega = \left((0, T)\times\openOmega\right)\cup\Gamma_1^0\cup\Gamma_2^0\cup\Gamma_1^+\cup\Gamma_2^+\cup \Bigl(\{0\}\times\Omega\Bigr),
        \end{equation*}
        by a uniform discretization of each of the resulting subsets of the decomposition, although for illustrative purposes we present the simplest case. We denote as $N_{S_1},\, N_{S_2}$ and $N_{T}$ the number of steps in the $S_1$, $S_2$ and time-direction. Given these values, the grids are given by
        \begin{align*}
             \mathcal{P}_{\I}&=\bigl\{(t_i, S_{1, j}, S_{2, k})\bigr\}_{i, j, k = 1}^{N_T, N_{S_1}-1, N_{S_2}-1} = \bigl\{\left(i\,\Delta_T, j\,\Delta_{S_1}, k\,\Delta_{S_2}\right):\, i, j, k=1,\dots N_T,N_{S_1}-1, N_{S_2}-1\bigr\}, \\
             \mathcal{P}_{\Gamma_1^0} &= \bigl\{(t_i, 0, S_{2, k})\bigr\}_{i=1, k=0}^{N_T, N_{S_2}} = \bigl\{(i\,\Delta_T, 0, k\,\Delta_{S_2})\, : \, i = 1,\dots N_T,\,k = 0 \dots N_{S_2}\bigr\},\\
             \mathcal{P}_{\Gamma_2^0} &= \bigl\{(t_i, S_{1,j}, 0)\bigr\}_{i, j = 1}^{N_T, N_{S_1}} = \bigl\{(i\,\Delta_T, j\,\Delta_{S_1}, 0)\,:\,i, j = 1,\dots N_T, N_{S_1}\bigr\},\\
             \mathcal{P}_{\Gamma_1^+} &= \bigl\{(t_i, S_{1, max}, S_{2, k})\bigr\}_{i, k = 1}^{N_T, N_{S_2}-1} = \bigl\{(i\,\Delta_T, S_{1, max}, k\,\Delta_{S_2})\,:\,i, k = 1, \dots N_T, N_{S_2}-1\bigr\},\\
             \mathcal{P}_{\Gamma_2^+} &= \bigl\{(t_i, S_{1, j}, S_{2, max})\bigr\}_{i, j=1}^{N_T, N_{S_1}} = \bigl\{(i\,\Delta_T, j\,\Delta_{S_1}, S_{2, max})\,:\,i, j = 1, \dots N_T, N_{S_1}\bigr\},\\
             \mathcal{P}_{\OO}&= \bigl\{(0, S_{1, j}, S_{2, k})\bigr\}_{j, k = 0}^{N_{S_1},\,\, N_{S_2}} = \bigl\{\left(0, j\,\Delta_{S_1}, k\,\Delta_{S_2}\right) : \, j, k=0, \dots N_{S_1}, N_{S_2}\bigr\},
        \end{align*}
        with $\Delta_{S_1} = S_{1, max} / N_{S_1},\, \Delta_{S_2} = S_{2, max} / N_{S_2}$ and $\Delta_T = T / N_T$ the step size related to each cartesian direction. Thus, the set of collocation points is given by
        \begin{equation*}
            \mathcal{P} = \mathcal{P}_{\I}\cup \mathcal{P}_{\Gamma_1^0} \cup \mathcal{P}_{\Gamma_2^0} \cup \mathcal{P}_{\Gamma_1^+} \cup \mathcal{P}_{\Gamma_2^+}\cup \mathcal{P}_{\OO},
        \end{equation*}
        with size
        \begin{equation*}
            N_{\mathcal{P}} = |\mathcal{P}| = (N_T + 1)\times (N_{S_1} + 1) \times (N_{S_2} + 1).
        \end{equation*}
        
        In order to obtain an approximate solution to the problems, a neural network $\hat{V_{\theta}}: [0, T]\times\Omega\subset\R^3 \longrightarrow \R$ under the same structural assumptions as for the one-dimensional case is considered. For both problems we can take the same residuals, except to that related to the initial condition, following the strategy presented in Section \ref{sec:trainingStrategy}, so that, 
        \begin{align} \label{eq:BS2D_res_1}
            \begin{split}
                \mathcal{R}_{\theta}^{\mathcal{I}} &=
                \dfrac{\partial\hat{V}_{\theta}}{\partial t} - \dfrac{\sigma_1^2 S_1^2}{2}\dfrac{\partial^2\hat{V}_{\theta}}{\partial S_1^2} - 
                \rho\sigma_1\sigma_2 S_1 S_2\dfrac{\partial^2\hat{V}_{\theta}}{\partial S_1 \partial S_2}
                - \dfrac{\sigma_2^2 S_2^2}{2}\dfrac{\partial^2\hat{V}_{\theta}}{\partial S_2^2}\\
                &- r_{R_1}S_1\dfrac{\partial\hat{V}_{\theta}}{\partial S_1} - r_{R_2} S_2 \dfrac{\partial\hat{V}_{\theta}}{\partial S_2} + r\hat{V}_{\theta} + f(\hat{V}_{\theta}),
            \end{split}
             & \text{in}&\,\, (0, T)\times\openOmega, \\ 
             \label{eq:BS2D_res_2}
            \mathcal{R}_{\theta}^{\Gamma_i^0} &= \dfrac{\partial\hat{V}_{\theta}}{\partial t} - \dfrac{\sigma_j^2 S_j^2}{2}\dfrac{\partial^2\hat{V}_{\theta}}{\partial S_j^2} - r_{R_j} S_j \dfrac{\partial \hat{V}_{\theta}}{\partial S_j} + r\hat{V}_{\theta} + f(\hat{V}_{\theta}), & \text{in} &\,\, \Gamma_i^0,\quad i=1, 2,\,j\neq i,\\ 
            \label{eq:BS2D_res_3}
            \begin{split}
                \mathcal{R}_{\theta}^{\Gamma_i^+} &= 
                \dfrac{\partial\hat{V}_{\theta}}{\partial t} - \rho\sigma_i\sigma_j S_i S_j\dfrac{\partial^2\hat{V}_{\theta}}{\partial S_i \partial S_j}
                - \dfrac{\sigma_j^2 S_j^2}{2}\dfrac{\partial^2\hat{V}_{\theta}}{\partial S_j^2} - r_{R_i}S_i\dfrac{\partial\hat{V}_{\theta}}{\partial S_i}  \\
                &- r_{R_j} S_j \dfrac{\partial\hat{V}_{\theta}}{\partial S_j} + r\hat{V}_{\theta} + f(\hat{V}_{\theta}),    
            \end{split}
             & \text{in} &\,\,\Gamma_i^+,\quad i=1, 2,\,j\neq i,\\
            \label{eq:BS2D_res_4}
            \mathcal{R_{\theta}^O} &= \hat{V}_{\theta} - H, & \text{in} &\,\, \{0\}\times\Omega,
        \end{align}
        where $H$ is given by \eqref{eq:average_payoff} or \eqref{eq:worst_of_payoff} for the arithmetic average or the worst-of option, respectively . Taking such residuals into account, it is straightforward to obtain an expression for the loss function similar to \eqref{eq:lossBS1d}.
    
    \subsubsection{European option under the Heston model} \label{sec:Heston}
        The last problem we address is the pricing of a European option accounting for CCR, with strike $K\in\R$ and maturity $T>0$, under the assumption that the variance of the underlying follows a stochastic process. Thus, let $S$ be the underlying stock value and $r_R$ the stock repo rate minus the dividend yield. We define the volatility of $S$ from its variance, $\nu$, which follows a CIR process, \cite{cir}, with $\eta>0$ the mean variance, $\kappa>0$ the mean reversion rate, $\sigma>0$ the volatility of the variance and $\rho\in[-1, 1]$ the correlation between the asset and variance processes. Under this setting, the Heston model is obtained, \cite{heston1993closed}.
        
        The PDE problem for pricing the risky European option under the Heston model is derived in \cite{salvador2021total}. The option price $\hat{V}$ is the solution of the equation \eqref{GeneralRiskyPDE} taking the elliptic operator
        \begin{equation} \label{eq:HestonOperator}
            \mathcal{L} = -\dfrac{S^2\nu}{2}\dfrac{\partial^2}{\partial S^2} - \rho\sigma S \nu \dfrac{\partial^2}{\partial S\partial \nu} - \dfrac{\sigma^2\nu}{2}\dfrac{\partial^2}{\partial \nu^2} - r_R S \dfrac{\partial}{\partial S} - \kappa (\eta - \nu)\dfrac{\partial}{\partial \nu} + r\mathcal{I},
        \end{equation}
        and as an initial condition the vanilla payoff \eqref{vanilla_payoff}.
        
        As in the previous case, it is necessary to establish an effective domain in order to apply numerical methods. Thus, we define our computational domain as $\Omega = [0, S_{max}]\times [0, \nu_{max}]$ and , again, additional conditions must be imposed over the boundaries  $\Gamma_S^0 = (0, T)\times\{0\}\times[0, \nu_{max}]$, $\Gamma_{\nu}^0 = (0, T)\times (0, S_{max}]\times\{0\}$, $\Gamma_S^+ = (0, T)\times\{S_{max}\}\times(0, \nu_{max})$ and $\Gamma_{\nu}^+ = (0, T)\times (0, S_{max}]\times\{\nu_{max}\}$.
        
        Following the boundary condition analysis carried out in \cite{salvador2021total} and \cite{CASTILLO}, it is not necessary to impose an additional condition on the boundary $\Gamma_S^0$. In addition, it will be only necessary to impose a condition on the boundary $\Gamma_{\nu}^0$ if the Feller condition, $2\kappa\eta>\sigma^2$, is violated. In such case, a common choice is to impose a Dirichlet condition obtained from the numerical resolution of the equation
        \begin{equation} \label{eq:feller_violated}
            \dfrac{\partial \hat{V}}{\partial t} - r_R S\dfrac{\partial\hat{V}}{\partial S} - \kappa\eta\dfrac{\partial\hat{V}}{\partial\nu} + r\hat{V} + f(\hat{V}) = 0,\qquad \text{in}\,\,\Gamma_{\nu}^0.
        \end{equation}
        On the boundary $\Gamma_S^+$ we keep the linearity condition \eqref{eq:BSRightBoundary}, while on the boundary $\Gamma_{\nu}^+$ we choose to employ the Neumann condition derived from the fact that
        \begin{equation} \label{hestonNeumannV}
            \lim_{\nu\to\infty} \dfrac{\partial \hat{V}}{\partial \nu}(t, S, \nu) = 0.
        \end{equation}
        We are in position to present the boundary value problem for pricing the risky European option under the Heston model. Therefore, we find $\hat{V}: [0, T] \times \Omega\subset\R^3 \longrightarrow \mathbb{R}$ such that
        \begin{equation} \label{eq:HestonBVP}
            \begin{cases}
                \!\begin{aligned}
                    \dfrac{\partial \hat{V}}{\partial t} - \dfrac{S^2\nu}{2}\dfrac{\partial^2\hat{V}}{\partial S^2} -& \rho\sigma S\nu\dfrac{\partial^2\hat{V}}{\partial S\partial\nu}-\dfrac{\sigma^2\nu}{2}\dfrac{\partial^2\hat{V}}{\partial \nu^2} \\-& r_R S\dfrac{\partial\hat{V}}{\partial S}
                    - \kappa(\eta - \nu)\dfrac{\partial\hat{V}}{\partial \nu} + r\hat{V}+f(\hat{V}) = 0,
                \end{aligned}
                 & \text{in}\,\,(0, T)\times \openOmega,\\
                \dfrac{\partial^2\hat{V}}{\partial S^2} = 0, & \text{in}\,\, \Gamma_S^+,\\
                \dfrac{\partial\hat{V}}{\partial\nu} = 0, & \text{in}\,\, \Gamma_{\nu}^+,\\
                \hat{V} - \max \Bigl\{\alpha \left(S - K\right), 0 \,\Bigr\} = 0, & \text{in}\,\,\{0\}\times\Omega,
            \end{cases}
        \end{equation}
        when the Feller condition is satisfied. Again, the risk-free Heston boundary problem is recovered by taking the risk parameters $\lambda_B = \lambda_C = 0$; and such formulations fits into the problem \eqref{EQ_parabolic}, so the techniques in Section \ref{sec:pinns} can be readily applied.
        
        At the methodological level, the development of this two-dimensional problem is similar to the one already seen. Starting from a discretization of the domain (we can think of the one given before), we define the residuals used in the training of a neural network $\hat{V}_{\theta}:[0, T]\times\Omega\subset\R^3 \longrightarrow \R$ in the task of approximating the solution of \eqref{eq:HestonBVP} as
        \begin{align} 
            \label{eq:ResHeston_I}
            \begin{split}
                \mathcal{R}_{\theta}^{\mathcal{I}} &= \dfrac{\partial \hat{V}_{\theta}}{\partial t} - \dfrac{S^2\nu}{2}\dfrac{\partial^2\hat{V}_{\theta}}{\partial S^2} - \rho\sigma S\nu\dfrac{\partial^2\hat{V}_{\theta}}{\partial S\partial\nu}\\
                &-\dfrac{\sigma^2\nu}{2}\dfrac{\partial^2\hat{V}_{\theta}}{\partial \nu^2}- r_R S\dfrac{\partial\hat{V}_{\theta}}{\partial S} - \kappa(\eta - \nu)\dfrac{\partial\hat{V}_{\theta}}{\partial \nu} + r\hat{V}_{\theta}+f(\hat{V}_{\theta}),    
            \end{split}
            & \text{in} &\,\,(0, T)\times\openOmega,\\
            \mathcal{R}_{\theta}^{\Gamma_{S}^+} &= \dfrac{\partial \hat{V}_{\theta}}{\partial t} - \rho\sigma S\nu\dfrac{\partial^2\hat{V}_{\theta}}{\partial S\partial\nu}-\dfrac{\sigma^2\nu}{2}\dfrac{\partial^2\hat{V}_{\theta}}{\partial \nu^2}- r_R S\dfrac{\partial\hat{V}_{\theta}}{\partial S} - \kappa(\eta - \nu)\dfrac{\partial\hat{V}_{\theta}}{\partial \nu} + r\hat{V}_{\theta}+f(\hat{V}_{\theta}),& \text{in} &\,\,\Gamma_{\nu}^+,\\ 
            \label{eq:ResHestonS0}
            \mathcal{R}_{\theta}^{\Gamma_{S}^0} &= \dfrac{\partial \hat{V}_{\theta}}{\partial t}  -\dfrac{\sigma^2\nu}{2}\dfrac{\partial^2\hat{V}_{\theta}}{\partial \nu^2} - \kappa(\eta - \nu)\dfrac{\partial\hat{V}_{\theta}}{\partial \nu} + r\hat{V}_{\theta}+f(\hat{V}_{\theta}),& \text{in} &\,\,\Gamma_{S}^0,\\ \label{eq:ResHestonV0}
            \mathcal{R}_{\theta}^{\Gamma_{\nu}^0} &= \dfrac{\partial \hat{V}_{\theta}}{\partial t} - r_R S\dfrac{\partial\hat{V}_{\theta}}{\partial S} - \kappa\eta\dfrac{\partial\hat{V}_{\theta}}{\partial\nu} + r\hat{V}_{\theta} + f(\hat{V}_{\theta}),& \text{in} &\,\,\Gamma_{\nu}^0,\\ \label{eq:resHeston_IC}
            \mathcal{R}_{\theta}^{\mathcal{O}} &= \hat{V_{\theta}} - \max \Bigl\{\alpha \left(S - K\right), 0 \,\Bigr\}, & \text{in} &\,\,\{0\}\times \Omega.
        \end{align}
        In this case, we decide to include the boundary-related residuals \eqref{eq:ResHestonS0} and \eqref{eq:ResHestonV0} as if they were boundary conditions, but they could be also considered as part of the interior of the domain straightforwardly. Then, $\hat{V}_{\theta}$ is trained by means of a loss function like the one presented in \eqref{eq:lossBS1d}, adapted to the residuals and higher dimension present here. 
    
    \section{Numerical experiments} \label{sec:experiments}
    After presenting the mathematical models and discussing how they fit under our reformulation via PINNs, in this section  we show the results of the tests performed to assess their effectiveness. One of the main advantages of this methodology over traditional numerical methods is that the container of the approximate solution is an ANN, i.e., a function. Thus, it is possible to compute its derivatives via AD. In this regard, we will focus not only on how well it approximates the desired solution, but also on how accurately it approximates its derivatives.

\begin{table}[!h]
    \small
    \centering
        \begin{tabular}{p{7cm}c}
    \hline 
    \multicolumn{2}{c}{Black Scholes parameters}                  \\
    \hline \hline
    Strike, K                               & $15$                  \\
    \hline
    Time to maturity, T                     & $5$                   \\
    \hline
    Volatility, $\sigma$                    & $0.25$                \\
    \hline
    Repo rate minus dividend, $r_R$         & $0.015$               \\
    \hline
    Interest rate, $r$                      & $0.03$                \\
    \hline \hline
    \multicolumn{2}{c}{xVA parameters}                            \\
    \hline 
    Seller hazard rate, $\lambda_B$         & $[0.0, 0.1]$                \\
    \hline
    Counterparty hazard rate, $\lambda_C$   & $0.05$                \\
    \hline
    Seller recovery rate, $R_B$             & $0.4$                 \\
    \hline
    Counterparty recovery rate, $R_C$       & $0.4$                 \\
    \hline 
    Funding spread, $s_F$                   & $(1 - R_B)\lambda_B$  \\
    \hline
\end{tabular}
    \caption{Parameters for Black-Scholes model considering counterparty risk, obtained from \cite{chen2019penalty}.}
    \label{tab:BS1d_parameters}
\end{table}

The section is divided into two parts. In the first, we focus on the one-dimensional parabolic problem, i.e., the pricing of options via Black-Scholes model; while the second covers two-dimensional parabolic problems, i.e., basket options and Heston option pricing. The same pattern is followed in both parts. First, an optimal network configuration, namely, the optimal number of layers and units per layer, is determined. For this purpose, the training metrics and the time required are taken into account. Subsequently, the error of the approximations is analyzed and, finally, tests relative to the computation of derivatives are presented. The reference values are computed by using the available analytic solutions or extremely reliable approximations based on classical resolution techniques such as FD or FE.

For the training, we use expression \eqref{DiscretizedLossFunctionVols} as the loss function, adapted by following the analysis carried out in Section \ref{sec:pdes} and taking all lambda weights equals to one. In addition, we choose the trapezoidal rule as the quadrature method. Consequently, we take an uniform grid of collocation points with variable size depending on the problem. Each training is split into two stages, depending on the employed optimizer. In the first stage, Adam is used as a global optimizer with the reference parameters given in \cite{kingma2014adam}, and, in the second one, L-BFGS is used as a local optimizer.

\subsection{Parabolic one-dimensional case}
    We study the one-dimensional parabolic case by means of the Black-Scholes equation presented in Section \ref{subsec:blackScholes}. For this purpose, we consider the model data presented in Table \ref{tab:BS1d_parameters} and choose $S_{max} = 4K$ as the truncation value of the domain. In addition, we work with a spatial discretization of $N_S=110$ points and a temporal discretization of $N_T=100$ points, yielding a total of $11,000$ collocation points, which falls within the reference values that can be found in other works, such as \cite{raissi2017physics}.
    
    First, a test is conducted to check how the network's training behaves when varying its number of layers and neurons per layer. For this purpose, all $16$ possible combinations between $l\in \{2, 4, 8, 16\}$ layers and $\beta\in \{10, 20, 40, 80\}$ units per layer are considered. For each combination, a sample of $10$ training is made. The pricing of an European put option, $V$, is the target, so we use the loss function \eqref{eq:lossBS1d} taking $\lambda_B = \lambda_C = 0$. The optimization process has $10,000$ steps with Adam and $2,500$ with L-BFGS. The accuracy of the PINNs solution is measured by comparing its relative error with the analytic solution \eqref{eq:BSFormula}.
    
    \begin{table}[!h]
        \begin{subtable}[t]{0.46\textwidth}
            \footnotesize
            \centering
            \begin{tabular}{|c||c|c|c|c|}
    \hline
    \backslashbox{Units}{Layers}        & 2             & 4                 & 8             & 16    \\
    \hline\hline 
                10                      &-3.664         &-3.531             &-3.669         &-0.149 \\
    \hline
                20                      &-3.182         &-3.344             &-3.465         &-2.99  \\
    \hline
                40                      &-3.357         &\textbf{-3.557}    &-3.301         &-0.159 \\
    \hline 
                80                      &-3.457         &-3.519             &-3.398         &-0.148 \\
    \hline
\end{tabular}
            \caption{Relative $L^1$ error in $\log$ scale.}
            \label{tab:1dCapacityL1}
        \end{subtable}
        \hfill
        \begin{subtable}[t]{0.46\textwidth}
            \footnotesize
            \centering
            \begin{tabular}{|c||c|c|c|c|}
    \hline
    \backslashbox{Units}{Layers}        & 2             & 4                 & 8             & 16    \\
    \hline\hline
                10                      &-3.113         &-3.098             &-0.937         &-0.202 \\
    \hline
                20                      &-3.160         &-3.308             &-3.396         &-2.969 \\
    \hline
                40                      &-3.351         &\textbf{-3.447}    &-3.356         &-0.188 \\
    \hline
                80                      &-3.382         &-3.441             &-3.419         &-0.209 \\
    \hline
\end{tabular}
            \caption{Relative $L^2$ error in $\log$ scale.}
            \label{tab:1dCapacityL2}
        \end{subtable}
        \hfill
        \begin{subtable}[t]{0.46\textwidth}
            \footnotesize
            \centering
            \begin{tabular}{|c||c|c|c|c|}
    \hline
    \backslashbox{Units}{Layers}        & 2             & 4                 & 8             & 16    \\
    \hline\hline
                10                      &-2.893         &-2.986             &-0.855         &-0.214 \\
    \hline
                20                      &-3.001         &-3.134             &-3.216         &-2.852 \\
    \hline
                40                      &-3.177         &\textbf{-3.206}    &-3.160         &-0.152 \\
    \hline
                80                      &-3.190         &-3.207             &-3.226         &-0.189 \\
    \hline
\end{tabular}
            \caption{Relative $L^{\infty}$ error in $\log$ scale.}
            \label{tab:1dCapacityLinf}
        \end{subtable}
        \hfill
        \begin{subtable}[t]{0.46\textwidth}
            \footnotesize
            \centering
            \begin{tabular}{|c||c|c|c|c|}
    \hline
    \backslashbox{Units}{Layers}        & 2             & 4                 & 8             & 16    \\
    \hline\hline
                10                      &0.465          &0.536              &0.677          &0.721  \\
    \hline
                20                      &0.468          &0.547              &0.693          &0.928  \\
    \hline
                40                      &0.481          &\textbf{0.590}     &0.758          &1.000  \\
    \hline
                80                      &0.530          &0.680              &0.946          &0.926  \\
    \hline
\end{tabular}
            \caption{Relative training time.}
            \label{tab:1dCapacityTime}
        \end{subtable}
        \label{tab:1dCapacity}
        \caption{Worst relative error and training time achieved for each combination of layers and neurons per layer in the one-dimensional case.}
    \end{table}
    
    Tables \ref{tab:1dCapacityL1}, \ref{tab:1dCapacityL2} and \ref{tab:1dCapacityLinf} show the relative error achieved in the worst training for each considered combination of layers and units per layer. As we can see, most of the combinations give good results, with the $L^2$ relative error similar to those obtained for the same task in \cite{math_Bea}, where the tuning of the lambda weights is performed and the Monte Carlo integration is employed as a quadrature rule. From these tables we can also see that the use of a large number of layers is unstable, since the convergence of the method fails for some trials. This situation is possibly related to problems in updating the network's weights, such as vanishing gradient problems, due to the combination of very deep networks and bounded activation functions, \cite{geron}.
    
    Trying to find a balance between accuracy, robustness and performance, we have measured the training time for each studied combination and, in Table \ref{tab:1dCapacityTime}, we present the relative times obtained with respect to the largest one. Based on them, in what follows, we work with $l=4$ layers and $\beta=40$ units per layer, where we have achieved, in the best case, a $\log$ relative error of $-3.632$, $-3.538$ and $-3.290$ for the $L^1$, $L^2$ and $L^{\infty}$-norm, respectively.
    
    \begin{figure}[!h]
        \centering
        \includegraphics[width=0.65\textwidth, page=1]{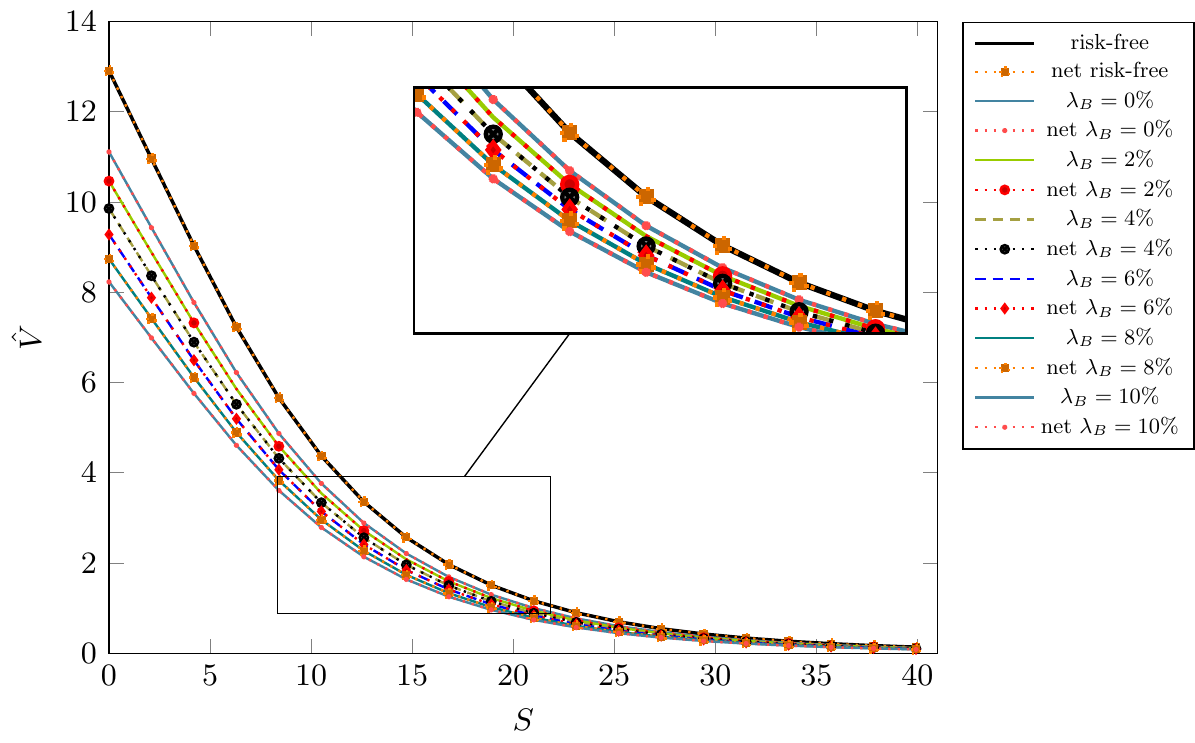}
        \caption{Comparison between analytical and approximated put option values for each default scenario.}
        \label{fig:sellerRatesBS1DPrices}
    \end{figure}
    
    \begin{figure}[!h]
        \centering
        \begin{subfigure}[b]{0.49\textwidth}
            \centering
            \includegraphics[width=\textwidth, page=2]{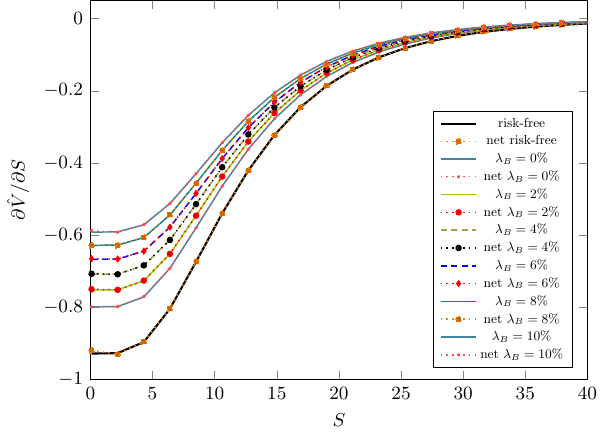}
            \caption{Put option deltas.}
            \label{fig:sellerRatesBS1DDeltas}
        \end{subfigure}
        \hfill
        \begin{subfigure}[b]{0.49\textwidth}
            \centering
            \includegraphics[width=\textwidth, page=3]{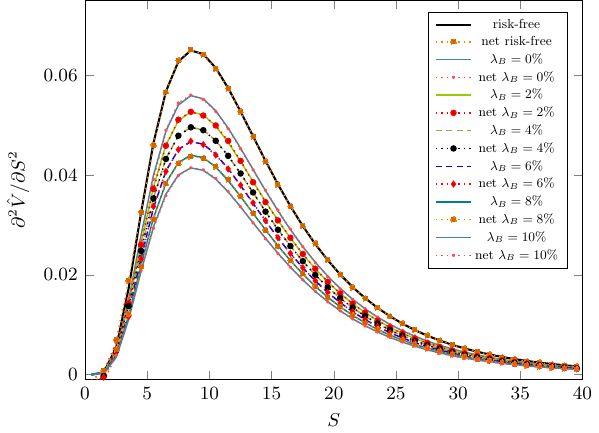}
            \caption{Put option gammas.}
            \label{fig:sellerRatesBS1DGammas}
        \end{subfigure}
        \caption{Comparison between analytical and approximated put option deltas and gammas for each default scenario.}
        \label{fig:sellerRatesGreeks}
    \end{figure}
    
    Once the size of our network has been selected, we present some results on its performance for the non-linear case. To do this, we consider six possible default scenarios depending on the seller hazard rate, namely $\lambda_B\in \{0\%, 2\%, 4\%, 6\%, 8\%, 10\%\}$, and train our network to price a risky put option with the rest of the parameters given in Table \ref{tab:BS1d_parameters}. As we did before, we perform $10$ samples per $\lambda_B$ and we select network weights that give the best result. The optimization process setting is kept invariant. The obtained solution is compared with the analytical solution \eqref{eq:riskyBSFormula}.
    
    Figure \ref{fig:sellerRatesBS1DPrices} shows the comparison between the analytical and PINNs approximated solution for each $\lambda_B$ considered. The risk-free option is added for completeness. Regardless of the default scenario chosen, 
    the quantitative behaviour of our approximation is identical to that given by the analytical solution. The accuracy of the approximation is particularly good in the neighbourhood of the strike, an area of interest in our pricing task. This is supported by Table \ref{tab:relError1D}, where we can see that, for all cases, the error is of the order of $10^{-4}$. Moreover, we observe that there does not exist any loss of accuracy in the non-linear cases, thus requiring no further treatment.
    
    \begin{table}[!h]
        \small
        \centering
            \begin{tabular}{c c|c|c|c}
    Case & $S$ &  $\hat{V}$ & $\partial \hat{V}/\partial S$ & $\partial^2 \hat{V}/\partial S^2$ \\
    \hline
    \multirow{3}{6em}{Risk-free}& $12.5$   & $5.55\times 10^{-4}$   & $8.87\times 10^{-5}$   & $1.39\times 10^{-4}$  \\
                                & $15.0$   & $6.99\times 10^{-4}$   & $4.51\times 10^{-4}$   & $2.22\times 10^{-3}$  \\ 
                                & $17.5$   & $6.71\times 10^{-4}$   & $8.82\times 10^{-4}$   & $2.36\times 10^{-3}$  \\
    \hline
    \multirow{3}{6em}{$\lambda_B = 0\%$}    & $12.5$   & $3.11\times 10^{-4}$   & $1.85\times 10^{-4}$   & $2.54\times 10^{-3}$   \\ 
                                             & $15.0$   & $2.81\times 10^{-4}$   & $2.87\times 10^{-4}$   & $3.30\times 10^{-3}$   \\ 
                                             & $17.5$   & $5.04\times 10^{-4}$   & $9.10\times 10^{-4}$   & $5.33\times 10^{-4}$   \\ 
    \hline
    \multirow{3}{6em}{$\lambda_B = 2\%$} & $12.5$   & $2.28\times 10^{-4}$   & $5.92\times 10^{-4}$   & $2.42\times 10^{-3}$   \\ 
                                         & $15.0$   & $1.79\times 10^{-4}$   & $2.60\times 10^{-4}$   & $2.57\times 10^{-3}$   \\ 
                                         & $17.5$   & $3.48\times 10^{-4}$   & $1.43\times 10^{-5}$   & $2.07\times 10^{-3}$   \\
    \hline
    \multirow{3}{6em}{$\lambda_B = 4\%$}  & $12.5$   & $1.27\times 10^{-4}$   & $1.49\times 10^{-4}$   & $1.84\times 10^{-3}$   \\ 
                                         & $15.0$   & $1.50\times 10^{-4}$   & $5.67\times 10^{-5}$   & $2.23\times 10^{-3}$   \\ 
                                         & $17.5$   & $2.98\times 10^{-4}$   & $3.63\times 10^{-4}$   & $1.11\times 10^{-3}$   \\ 
    \hline
    \multirow{3}{6em}{$\lambda_B = 6\%$}   & $12.5$   & $1.64\times 10^{-4}$   & $7.23\times 10^{-4}$   & $1.93\times 10^{-3}$   \\
                                            & $15.0$   & $3.76\times 10^{-4}$   & $4.84\times 10^{-5}$   & $1.45\times 10^{-3}$   \\ 
                                            & $17.5$   & $4.91\times 10^{-4}$   & $1.41\times 10^{-4}$   & $2.69\times 10^{-4}$   \\ 
    \hline
    \multirow{3}{6em}{$\lambda_B = 8\%$}  & $12.5$   & $6.48\times 10^{-4}$   & $3.21\times 10^{-4}$   & $2.62\times 10^{-3}$   \\
                                         & $15.0$   & $6.54\times 10^{-4}$   & $6.98\times 10^{-4}$   & $3.14\times 10^{-3}$   \\ 
                                         & $17.5$   & $7.58\times 10^{-4}$   & $3.75\times 10^{-4}$   & $2.02\times 10^{-3}$   \\ 
    \hline
    \multirow{3}{6em}{$\lambda_B = 10\%$}   & $12.5$   & $1.01\times 10^{-4}$   & $4.82\times 10^{-4}$   & $8.89\times 10^{-4}$   \\ 
                                             & $15.0$   & $1.67\times 10^{-5}$   & $1.17\times 10^{-4}$   & $2.40\times 10^{-3}$   \\ 
                                             & $17.5$   & $4.51\times 10^{-6}$   & $6.86\times 10^{-5}$   & $1.21\times 10^{-3}$   \\ 
    \hline
\end{tabular}

        \caption{Relative errors for the put option price, delta and gamma, with $S$ near the strike, for each default scenario. Risk-free case ($\lambda_B=\lambda_C=0$) is added for completeness.}
        \label{tab:relError1D}
    \end{table}
    
    By means of the AD, we can compute the derivative of the option price with respect to its related quantities. Such expressions are known as Greeks in quantitative finance. Thus, in Figure \ref{fig:sellerRatesBS1DDeltas} and Figure \ref{fig:sellerRatesBS1DGammas}, we can observe the same comparison made for the price, now for delta and gamma Greeks\footnote{Delta and gamma Greeks are, respectively, the first and second-order derivative of the option price with respect to its underlying.}, respectively. In the delta case, a slight decrease in accuracy is observed near the boundary $S=0$, which also transfers to the gamma case, as expected. In the rest of the domain there is not a loss of accuracy with respect to the pricing case. Specially in the neighbourhood of the strike, where we obtain relative errors of a similar order of magnitude, see Table \ref{tab:relError1D}. In the case of the second derivative we observe, in general, an increase in the relative error, now of the order of $10^{-3}$. This is also expected since it presents numerical instabilities that makes it more difficult to compute.

\subsection{Parabolic two-dimensional case}
    Having seen the results obtained for the one-dimensional Black-Scholes equation, with and without considering counterparty risk, we now present the results obtained for the rest of the presented models.
    \subsubsection{Basket options under the Black-Scholes model}
         We first deal with the basket options, whose formulation has been presented in Section \ref{subsec:basket}. For this purpose, we consider the model data given in Table \ref{tab:BS2d_params}, and choose as the truncation values of the domain $S_{1,max}=S_{2, max}=4K$.
        
        \begin{table}[!h]
        \small
            \centering
            \begin{tabular}{p{7cm}cc}
    \hline 
    \multicolumn{3}{c}{Black-Scholes parameters}                                             \\
    \hline \hline
    Strike, $K$                             & \multicolumn{2}{c}{$50$}                \\
    \hline
    Time to maturity, $T$                   & \multicolumn{2}{c}{$1$}                 \\
    \hline
    Interest rate, $r$                      &\multicolumn{2}{c}{$0.03$}               \\
    \hline \hline
                                            & $S_1$                         & $S_2$     \\
    \hline
    Volatility, $\sigma_i$                    & $0.25$                        & $0.15$    \\
    \hline
    Repo rate minus dividend, $r_{R_i}$         & $0.015$                       & $0.022$   \\
    \hline
    Correlation, $\rho$                     &\multicolumn{2}{c}{$-0.65$}              \\
    \hline \hline
    \multicolumn{3}{c}{xVA parameters}                                                \\
    \hline 
    Seller hazard rate, $\lambda_B$         & \multicolumn{2}{c}{$[0.0, 0.1]$}              \\
    \hline
    Counterparty hazard rate, $\lambda_C$   & \multicolumn{2}{c}{$0.07$}              \\
    \hline
    Seller recovery rate, $R_B$             & \multicolumn{2}{c}{$0.5$}               \\
    \hline
    Counterparty recovery rate, $R_C$       & \multicolumn{2}{c}{$0.3$}                \\
    \hline 
    Funding spread, $s_F$                   & \multicolumn{2}{c}{$(1 - R_B)\lambda_B$} \\
    \hline
\end{tabular}
            \caption{Parameters for the $2$-dimensional Black-Scholes model considering counterparty risk.}
            \label{tab:BS2d_params}
        \end{table}
        
        As in the previous case, we are interested in finding an optimal combination of layers and neurons per layer in terms of accuracy and training time required. Thus, we consider all $16$ possible combinations between  $l\in\{2, 4, 8, 12\}$ layers and $\beta\in\{10, 20, 40, 60\}$ units per layer. Again, a sample of $10$ training trials is considered per combination. We use the pricing of a non-risky arithmetic average put option as a target, so we use the loss function given by the residuals \eqref{eq:BS2D_res_1}-\eqref{eq:BS2D_res_4} with $\lambda_B = \lambda_C = 0$.
        
        In the training stage we use a total of $37,044$ collocation points ($N_T=21,\,N_{S_1}=N_{S_2}=42$), and the optimization process has $20,000$ steps with Adam and $2500$ with L-BFGS. Since the analytical solution for such options is not known, we measure the accuracy of the PINNs solution by comparing its relative error with an approximated solution of the non-risky boundary value problem \eqref{eq:RiskyAverageBasket} obtained via FD (Crank-Nicolson timestepping and centered differences).
        
        \begin{table}[ht]
            \begin{subtable}[t]{0.46\textwidth}
                \footnotesize
                \centering
                \begin{tabular}{|c||c|c|c|c|}
    \hline
    \backslashbox{Units}{Layers}    & 2         & 4                 & 8         & 12    \\
    \hline\hline
            10                      &-2.285     &-2.143             &-2.360     &-2.190 \\
    \hline
            20                      &-2.599     &-2.503             &-2.776     &-2.565 \\
    \hline
            40                      &-2.832     &-3.013             &-3.403     &-2.190 \\
    \hline
            60                      &-2.910     &\textbf{-3.430}    &-3.470     &-3.657 \\
    \hline
\end{tabular}

                \caption{Relative $L^1$ error in $\log$ scale.}
                \label{tab:2dCapacityL1}
            \end{subtable}
            \hfill
            \begin{subtable}[t]{0.46\textwidth}
                \footnotesize
                \centering
                \begin{tabular}{|c||c|c|c|c|}
    \hline
    \backslashbox{Units}{Layers}    & 2             & 4                 & 8         & 12    \\
    \hline\hline
            10                      &-2.127         &-2.067             &-2.233     &-2.088 \\
    \hline
            20                      &-2.466         &-2.380             &-2.705     &-2.473 \\
    \hline 
            40                      &-2.688         &-2.908             &-3.339     &-2.127 \\
    \hline
            60                      &-2.728         &\textbf{-3.326}    &-3.418     &-3.553 \\
    \hline
\end{tabular}
                \caption{Relative $L^2$ error in $\log$ scale.}
                \label{tab:2dCapacityL2}
            \end{subtable}
            \hfill
            \begin{subtable}[t]{0.46\textwidth}
                \footnotesize
                \centering
                \begin{tabular}{|c||c|c|c|c|}
    \hline
    \backslashbox{Units}{Layers}    & 2             & 4                 & 8             & 12    \\
    \hline
    \hline
                10                  &-1.785         &-1.702             &-1.971         &-1.626 \\
    \hline
                20                  &-2.194         &-2.117             &-2.399         &-2.059 \\
    \hline
                40                  &-2.308         &-2.530             &-2.886         &-1.928 \\
    \hline
                60                  &-2.328         &\textbf{-2.949}    &-2.907         &-2.976 \\
    \hline
\end{tabular}
                \caption{Relative $L^{\infty}$ error in $\log$ scale.}
                \label{tab:2dCapacityLinf}
            \end{subtable}
            \hfill
            \begin{subtable}[t]{0.46\textwidth}
                \footnotesize
                \centering
                \begin{tabular}{|c||c|c|c|c|}
    \hline
    \backslashbox{Units}{Layers}    & 2             & 4                 & 8             & 12    \\
    \hline\hline
                10                  &0.233          &0.291              &0.401          &0.994  \\
    \hline
                20                  &0.233          &0.291              &0.3400         &0.996  \\
    \hline
                40                  &0.232          &0.291              &0.400          &0.993  \\
    \hline
                60                  &0.233          &\textbf{0.288}     &0.392          &1.000  \\
    \hline
\end{tabular}
                \caption{Relative training time.}
                \label{tab:2dCapacityTime}
            \end{subtable}
            \label{tab:2dCapacity}
            \caption{Worst relative errors and training time achieved for each combination of layers and neurons per layer, considering the non-risky arithmetic average put option.}
        \end{table}
        
        Tables \ref{tab:2dCapacityL1}, \ref{tab:2dCapacityL2} and \ref{tab:2dCapacityLinf} show the relative error achieved in the worst training for each combination of layers and units per layer considered. We can observe a general increase in relative errors compared to the one-dimensional case, especially for combinations of layers and neurons that provide less capacity to the network. This is an expected situation since, on the one hand, the number of collocation points per spatial direction is much lower than in the previous case, and, on the other hand, the complexity of the function to be approximated increases. This situation is evident at-the-money\footnote{The at-the-money region is the subset of the underlyings' domain where the option's strike price is identical to the price given by the combination of the underlyings which defines the derivative contract. For example, the at-the-money region for the arithmetic average basket option is $\{(S_1, S_2)\in\Omega\,:\,S_1 + S_2 - 2K=0\}$. In this way, the out-the-money region is the domain's subset where the call (put) option's strike price is larger (smaller) than the price which defines the derivative contract, and the in-the-money region is its opposite.} (ATM), where there is a deterioration of the approximation in the presence of more complex payoff structures. We increase the number of collocation points in subsequent test to deal with this phenomenon. 
        
        In addition, we have also observed that the Adam's performance is suboptimal, in the sense that it comes to a point during training where it gets stuck. To avoid such situation, an adaptive learning rate is introduced in the following tests. We will use the so-called inverse time decay strategy, \cite{geron}, which follows the construction,
       \begin{equation*}
           \epsilon_{k} = \dfrac{\epsilon_0}{1 + \delta k /a},
       \end{equation*}
       where $\epsilon_0$ is the initial learning rate, $\epsilon_k$ the learning rate at step $k$, $\delta$ the decay rate and $a$ the decay step.
        
        \begin{figure}[!h]
            \centering
            \begin{subfigure}[b]{0.49\textwidth}
                \centering
                \includegraphics[width=\textwidth]{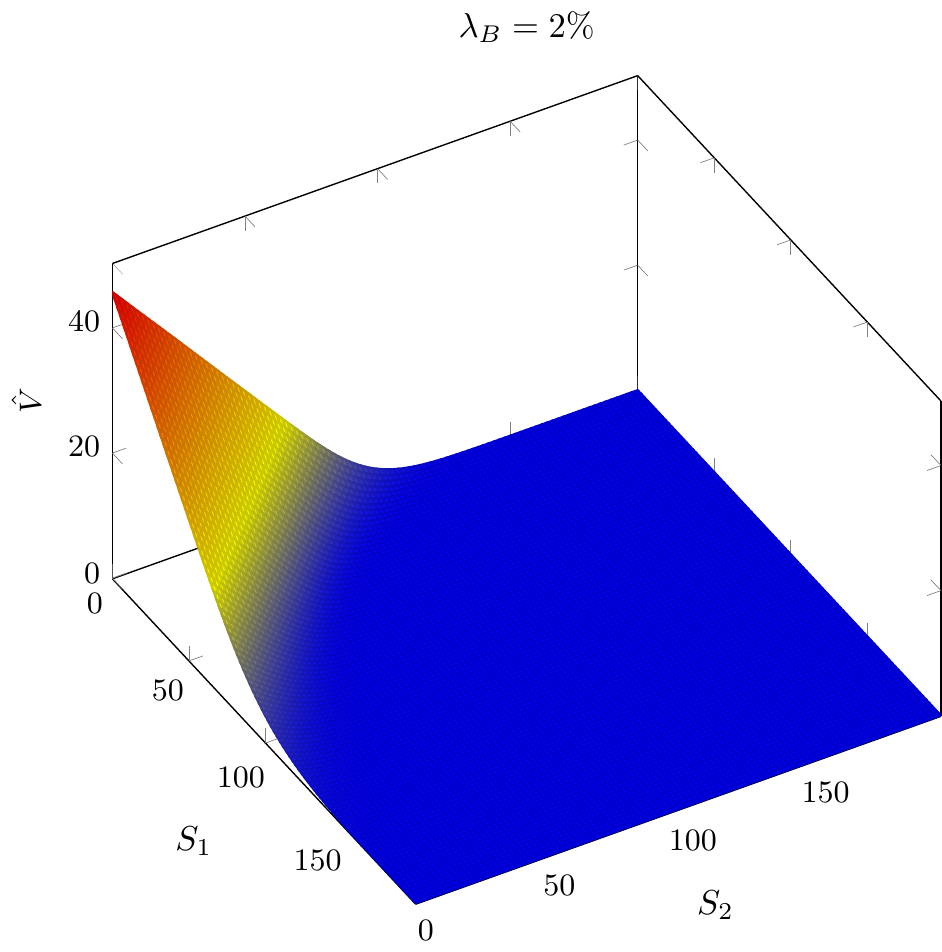}
                \caption{Put option value surface computed via trained ANN.}
                \label{fig:BasketSurface}
            \end{subfigure}
            \hfill
            \begin{subfigure}[b]{0.49\textwidth}
                \centering
                \includegraphics[width=\textwidth]{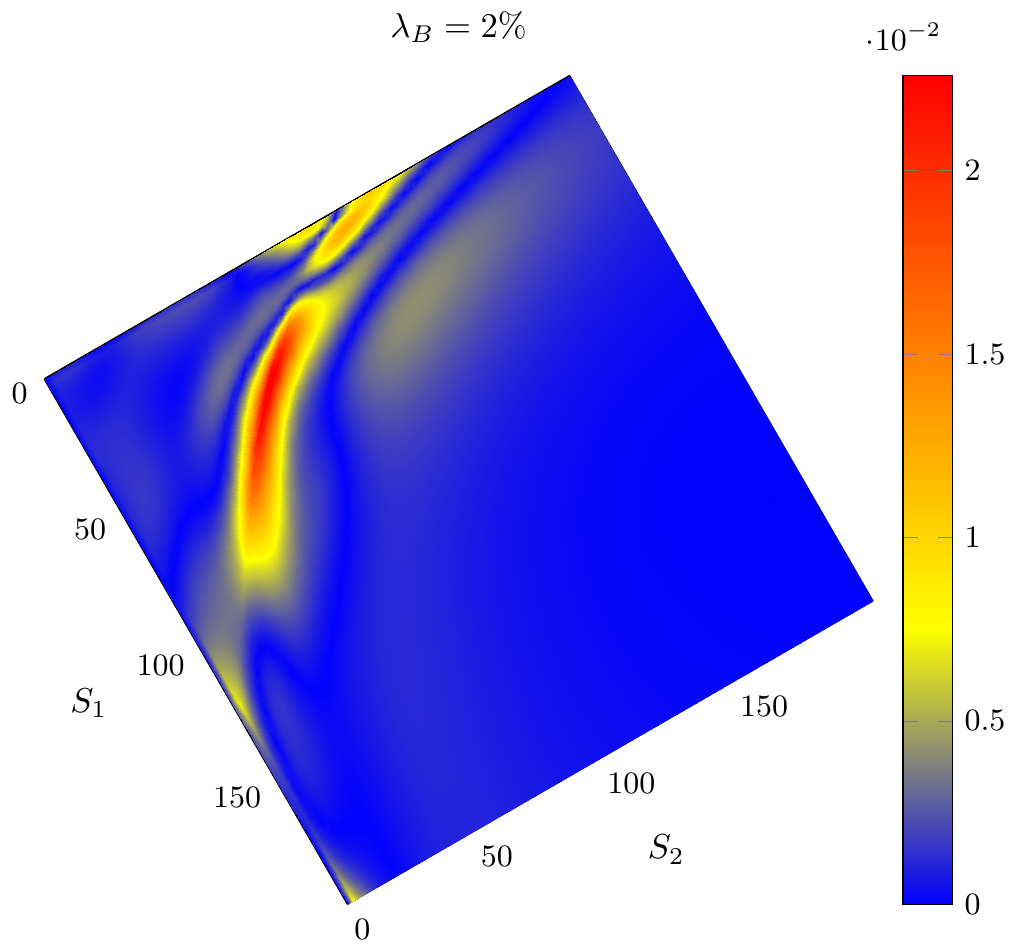}
                \caption{Relative error colour map.}
                \label{fig:BasketError}
            \end{subfigure}
            \caption{Risky arithmetic average put option with parameters given in Table \ref{tab:BS2d_params}.}
            \label{fig:BasketPlots}
        \end{figure}
        
        The increase in the problem dimension and in the number of input data also lead to a general increase in the time needed to perform the training, specially for the L-BFGS optimization step. Therefore, for the rest of the experiments in this work, we choose the combination of $l=4$ layers and $\beta=60$ units per layer, since it achieves relative errors very close to the best obtained, being about $30\%$ faster than the choices with less error, see Table \ref{tab:2dCapacityTime}. This setting has, in the best cases, a $\log$ relative error of $-3.518$, $-3.446$ and $-3.145$ for the $L^1$, $L^2$ and $L^{\infty}$ norms, respectively.
       
        In order to evaluate the performance of our training algorithm for the risky non-linear case, a test similar to the one performed in the one-dimensional case is run, now with the residuals given by \eqref{eq:BS2D_res_1}-\eqref{eq:BS2D_res_4}. For this purpose, we again consider six possible hazard rates, $\lambda_B\in \{0\%, 2\%, 4\%, 6\%, 8\%, 10\%\}$, being the remaining model parameters those given in Table \ref{tab:BS2d_params}. For each case, we do $5$ training trials with $141,204$ collocation points ($N_T=21,\,N_{S_1}=N_{S_2}=82$) and choose the best of them. We keep the number of Adam steps given before and take $\delta=0.75$, $a=5000$. The number of L-BFGS steps is also maintained.
        We take, as a reference, the solutions of the boundary value problem \eqref{eq:RiskyAverageBasket} obtained via FD with a fixed point scheme to deal with the non-linearity. Examples of this treatment can be found in, e.g., \cite{cva_pdes}, \cite{ARREGUI201731} or \cite{chen2019penalty}.
        
        \begin{figure}[!h]
            \centering
            \includegraphics[width=0.65\textwidth, page=1]{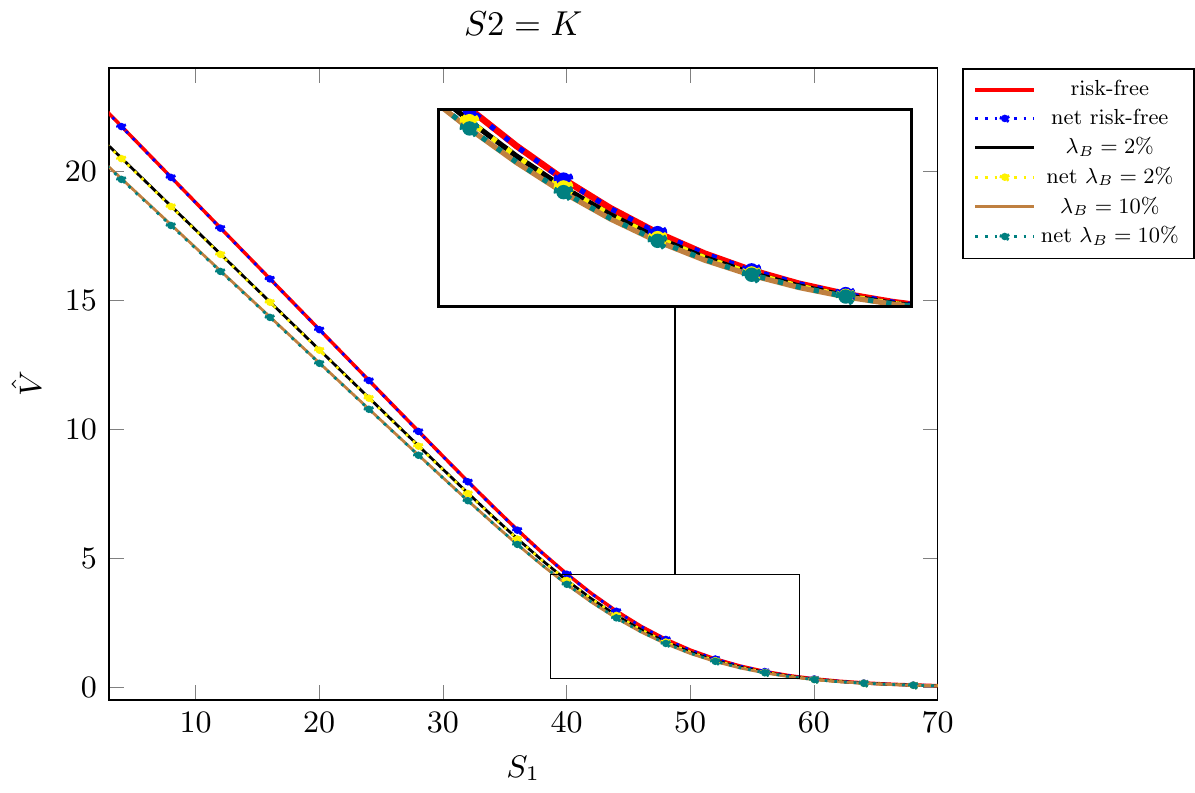}
            \caption{Comparison between finite differences and network arithmetic average put option values with $S_2=K$ fixed. Risk-free case ($\lambda_B=\lambda_C=0$), case $\lambda_B=2\%$ and case $\lambda_B=10\%$ are plotted.}
            \label{fig:sellerRatesBasketPrices}
        \end{figure}
        
        \begin{figure}[!h]
            \centering
            \begin{subfigure}[b]{0.49\textwidth}
                \centering
                \includegraphics[width=\textwidth, page=2]{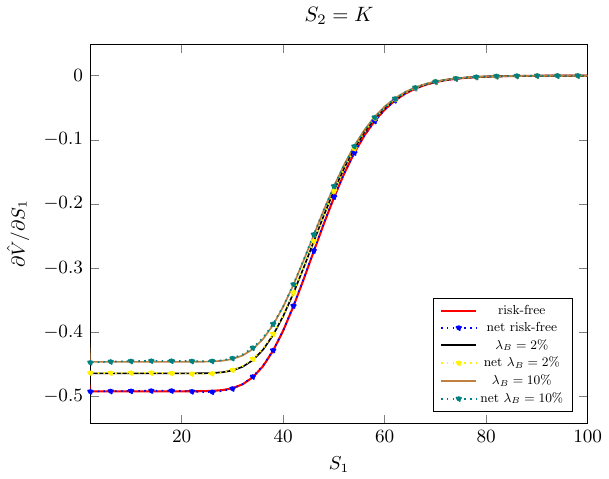}
                \caption{Delta with respect to the $S1$.}
                \label{fig:sellerRatesBasketDeltaS1}
            \end{subfigure}
            \hfill
            \begin{subfigure}[b]{0.49\textwidth}
                \centering
                \includegraphics[width=\textwidth, page=3]{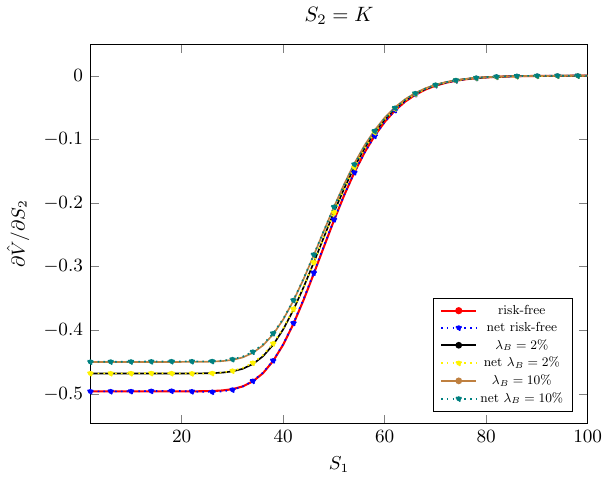}
                \caption{Delta with respect to the $S2$.}
                \label{fig:sellerRatesBasketDeltaS2}
            \end{subfigure}
            \caption{Comparison between finite differences and network arithmetic average put option deltas with $S_2=K$ fixed. Risk-free case ($\lambda_B=\lambda_C=0$), case $\lambda_B=2\%$ and case $\lambda_B=10\%$ are plotted.}
            \label{fig:sellerRatesBasketGreeks}
        \end{figure}
        
        \begin{table}[!h]
        \small
           \centering
           \begin{tabular}{c c|c|c|c}
    Case & $(S_1, S_2)$ &  $\hat{V}$ & $\partial \hat{V}/\partial S_1$ & $\partial\hat{V}/\partial S_2$ \\
    \hline
    \multirow{3}{6em}{Risk-free}    & $(50.0, 50.0)$  & $1.19\times 10^{-3}$    & $4.17\times 10^{-3}$    & $4.50\times 10^{-3}$       \\ 
                                    & $(42.9, 57.1)$  & $7.78\times 10^{-4}$    & $2.15\times 10^{-3}$    & $1.86\times 10^{-3}$       \\ 
                                    & $(57.1, 42.9)$  & $2.33\times 10^{-3}$    & $5.55\times 10^{-3}$    & $5.60\times 10^{-3}$       \\ 
                                    & $(55.0, 50.0)$  & $4.33\times 10^{-5}$    & $1.71\times 10^{-4}$    & $1.09\times 10^{-4}$       \\ 
                                    & $(50.0, 55.0)$  & $1.50\times 10^{-4}$    & $7.77\times 10^{-4}$    & $1.02\times 10^{-3}$     \\ 
                                    & $(55.0, 55.0)$  & $2.73\times 10^{-3}$    & $2.74\times 10^{-3}$    & $1.25\times 10^{-3}$      \\ 
                                    & $(45.0, 50.0)$  & $2.70\times 10^{-3}$    & $3.39\times 10^{-3}$    & $4.58\times 10^{-3}$       \\ 
                                    & $(50.0, 45.0)$  & $3.39\times 10^{-3}$    & $3.49\times 10^{-3}$    & $3.99\times 10^{-3}$      \\ 
                                    & $(45.0, 45.0)$  & $2.38\times 10^{-3}$    & $4.27\times 10^{-4}$    & $1.30\times 10^{-4}$      \\ 
    
    \hline
    \multirow{3}{6em}{$\lambda_B = 2\%$}    & $(50.0, 50.0)$ & $4.98\times 10^{-4}$    & $3.04\times 10^{-3}$    & $4.31\times 10^{-3}$      \\ 
                                            & $(42.9, 57.1)$ & $3.94\times 10^{-3}$    & $7.24\times 10^{-3}$    & $8.71\times 10^{-3}$       \\ 
                                            & $(57.1, 42.9)$ & $3.39\times 10^{-4}$    & $8.16\times 10^{-4}$    & $7.22\times 10^{-4}$       \\ 
                                            & $(55.0, 50.0)$ & $3.08\times 10^{-3}$    & $9.87\times 10^{-4}$    & $4.46\times 10^{-3}$       \\ 
                                            & $(50.0, 55.0)$ & $6.91\times 10^{-3}$    & $1.04\times 10^{-3}$    & $3.31\times 10^{-3}$       \\ 
                                            & $(55.0, 55.0)$ & $8.69\times 10^{-3}$    & $7.43\times 10^{-3}$    & $3.37\times 10^{-3}$       \\ 
                                            & $(45.0, 50.0)$ & $2.40\times 10^{-4}$    & $1.95\times 10^{-3}$    & $2.33\times 10^{-3}$       \\ 
                                            & $(50.0, 45.0)$ & $7.52\times 10^{-4}$    & $3.05\times 10^{-3}$    & $3.45\times 10^{-3}$       \\ 
                                            & $(45.0, 45.0)$ & $1.79\times 10^{-3}$    & $3.28\times 10^{-3}$    & $3.46\times 10^{-3}$       \\ 
                                        
    \hline
    \multirow{3}{6em}{$\lambda_B = 10\%$}   & $(50.0, 50.0)$ & $1.03\times 10^{-2}$    & $2.18\times 10^{-3}$    & $1.65\times 10^{-3}$   \\ 
                                            & $(42.9, 57.1)$ & $1.15\times 10^{-2}$    & $6.50\times 10^{-3}$    & $7.05\times 10^{-3}$    \\ 
                                            & $(57.1, 42.9)$ & $7.88\times 10^{-3}$    & $3.86\times 10^{-4}$    & $2.80\times 10^{-4}$     \\ 
                                            & $(55.0, 50.0)$ & $1.29\times 10^{-2}$    & $1.19\times 10^{-2}$    & $1.12\times 10^{-2}$     \\ 
                                            & $(50.0, 55.0)$ & $1.29\times 10^{-2}$    & $1.47\times 10^{-2}$    & $1.53\times 10^{-2}$     \\ 
                                            & $(55.0, 55.0)$ & $7.83\times 10^{-3}$    & $2.14\times 10^{-2}$    & $2.36\times 10^{-2}$    \\ 
                                            & $(45.0, 50.0)$ & $4.09\times 10^{-3}$    & $1.95\times 10^{-3}$    & $2.79\times 10^{-3}$    \\ 
                                            & $(50.0, 45.0)$ & $3.36\times 10^{-3}$    & $2.09\times 10^{-3}$    & $1.87\times 10^{-3}$    \\ 
                                            & $(45.0, 45.0)$ & $1.34\times 10^{-3}$    & $2.48\times 10^{-3}$    & $2.06\times 10^{-3}$   \\ 

    \hline
\end{tabular}
           \caption{Relative errors for the arithmetic average put option prices and deltas. The set of parameters given in Table \ref{tab:BS2d_params} is considered, taking $\lambda_B=\lambda_C=0$ in the risk-free case, and varying $\lambda_B$ in the others. For each case, three ATM, three out-the-money (OTM) and three in-the-money (ITM) values are taken.}
           \label{tab:relErrorBasket}
       \end{table}
        
        In Figure \ref{fig:BasketSurface} the PINNs solution for the risky arithmetic average put option, with $\lambda_B=2\%$, is plotted; while Figure \ref{fig:BasketError} shows the error compared to the reference solution. In order to avoid relative error instabilities due to values close to zero and thus obtain an adequate visualization of such error in the area of interest, its action is limited for option values greater than or equal to $0.01$. Regions with smaller option values are treated in terms of the absolute error, scaled by the imposed limit. This is sufficient for quantitative finance purposes and it is also followed in the error plots given below.
        
        As expected, the largest errors are observed ATM levels and its neighbourhood, as is the case in classical schemes. Even so, the errors in this region are reasonable, being at most of the order of $10^{-3}$. Figure \ref{fig:sellerRatesBasketPrices} shows a comparison between reference and network approximated prices in the same spirit as the one given for the previous case. We choose to show only two risky and the risk-free cases because the shorter maturity of the derivative leads to smaller adjustment between each scenario. In addition, slices of the first-order Greeks are added in Figure \ref{fig:sellerRatesBasketGreeks}. 
        
        It is observed that the solution approximated by the trained ANN has an identical qualitative behaviour in the plotted cases. In the remaining cases, not shown, there are no significant differences to comment on. For the first-order partial derivatives a similar behaviour to that given in the one-dimensional case is observed. They suffer from some slight oscillations near the lower boundaries, but show excellent results in the rest of the domain. In particular, we do not observe a decrease in the relative error compared to that obtained in prices. These facts are also supported by Table \ref{tab:relErrorBasket}, which shows the prices and Greeks' relative error for concrete combinations of $S_1$ and $S_2$.
       
       The same test is performed in the case of the worst-of put option, which is of interest in the industry because is a commonly offered product. We keep the test and parameters setup and we use the same methodology used in the arithmetic average case to compute the reference solutions.
        
        \begin{figure}[!h]
            \centering
            \begin{subfigure}[b]{0.49\textwidth}
                \centering
                \includegraphics[width=\textwidth]{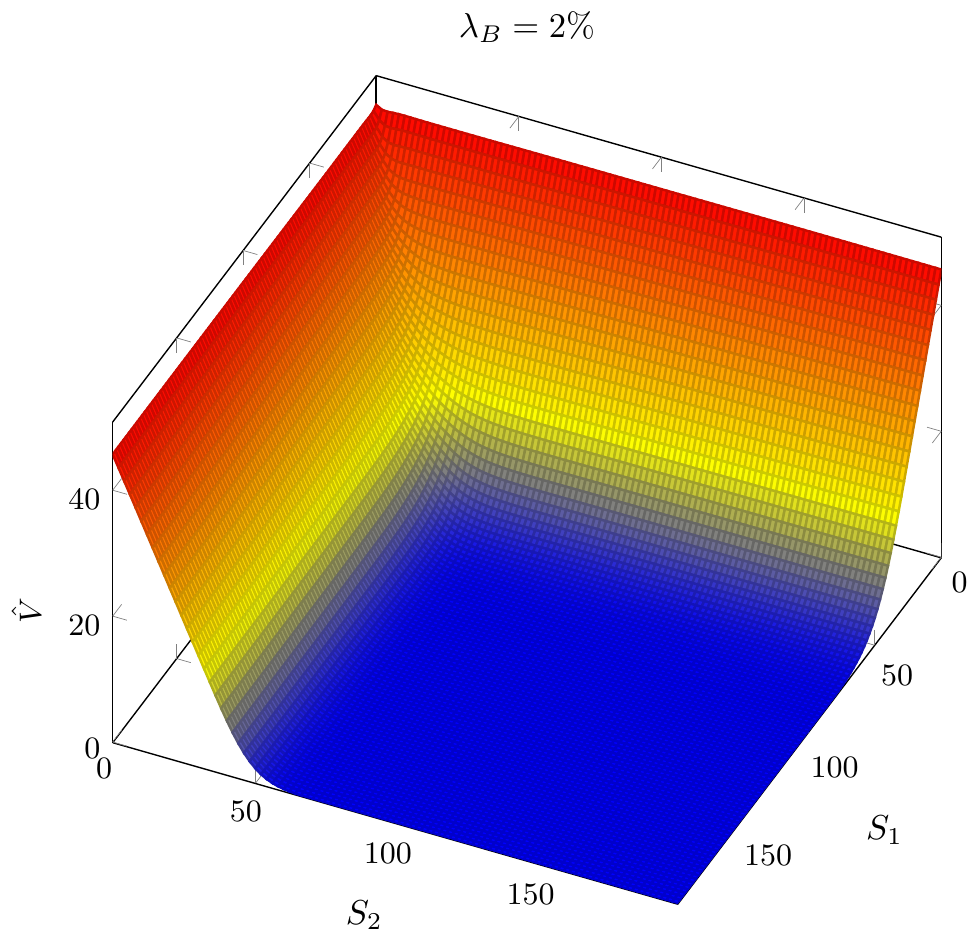}
                \caption{Put option value surface computed via trained ANN.}
                \label{fig:WorstOfSurface}
            \end{subfigure}
            \hfill
            \begin{subfigure}[b]{0.49\textwidth}
                \centering
                \includegraphics[width=\textwidth]{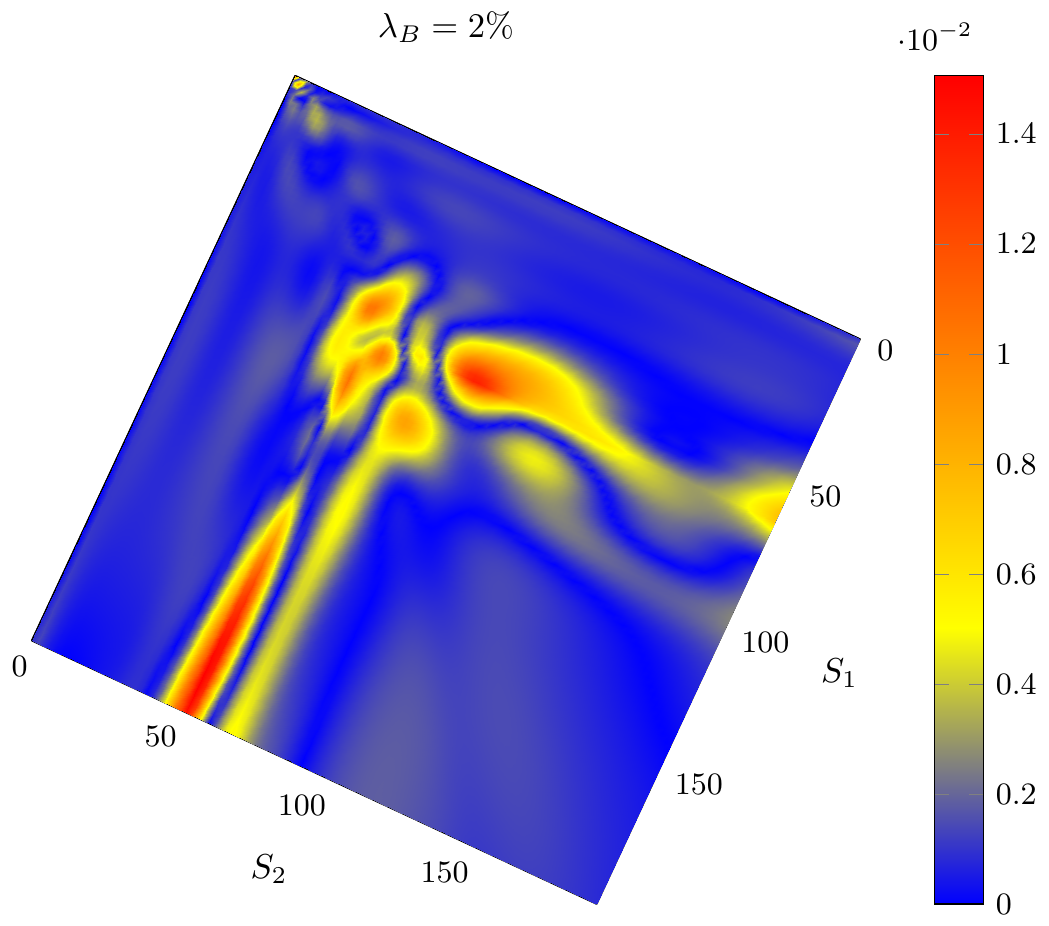}
                \caption{Relative error colour map.}
                \label{fig:WorstOfError}
            \end{subfigure}
            \caption{Risky worst-of put option with parameters given in Table \ref{tab:BS2d_params}.}
            \label{fig:WorstOfPlots}
        \end{figure}
        
        \begin{figure}[!h]
            \centering
            \includegraphics[width=0.65\textwidth, page=1]{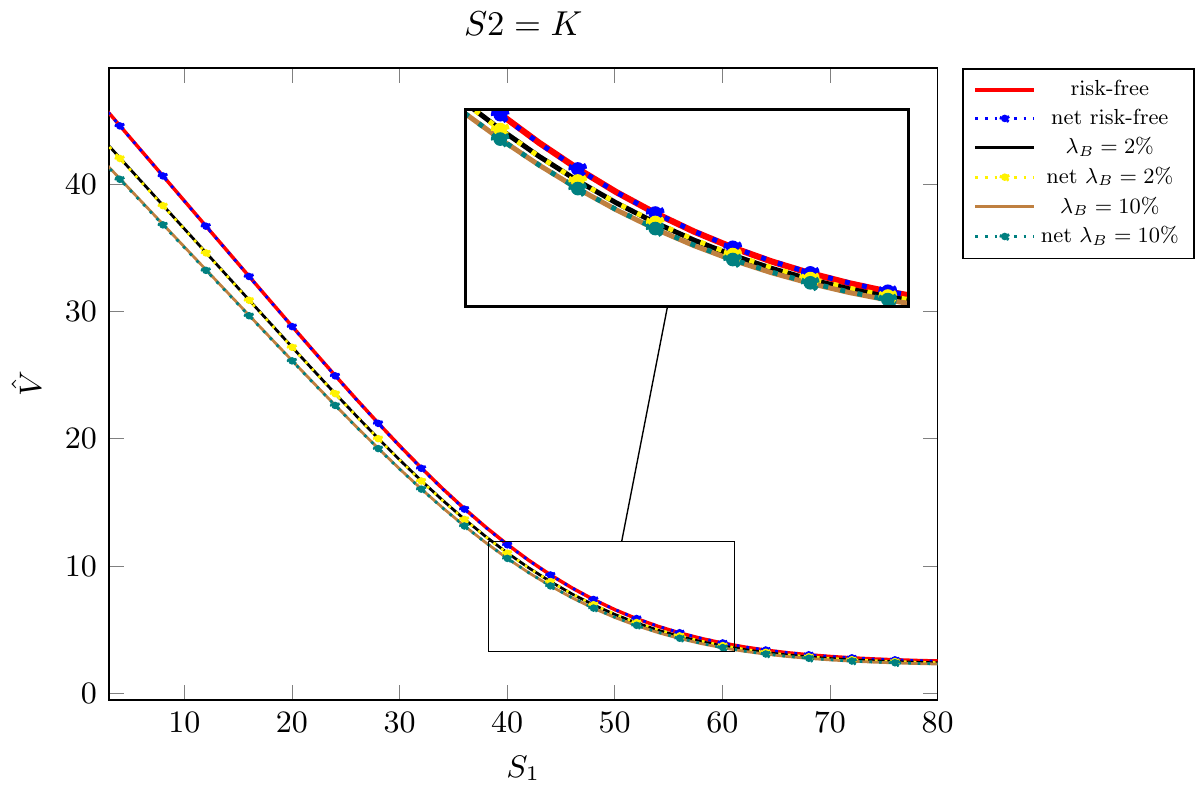}
            \caption{Comparison between finite differences and network worst-of put option values with $S_2=K$ fixed. Risk-free case ($\lambda_B=\lambda_C=0$), case $\lambda_B=2\%$ and case $\lambda_B=10\%$ are plotted.}
            \label{fig:sellerRatesWorstOfPrices}
        \end{figure}
        
        \begin{figure}[!h]
            \centering
            \begin{subfigure}[b]{0.49\textwidth}
                \centering
                \includegraphics[width=\textwidth, page=2]{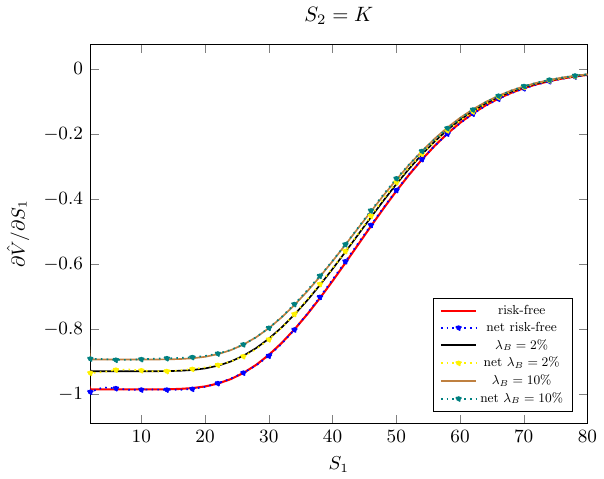}
                \caption{Delta with respect to the $S1$.}
                \label{fig:sellerRatesWorstOfDeltaS1}
            \end{subfigure}
            \hfill
            \begin{subfigure}[b]{0.49\textwidth}
                \centering
                \includegraphics[width=\textwidth, page=3]{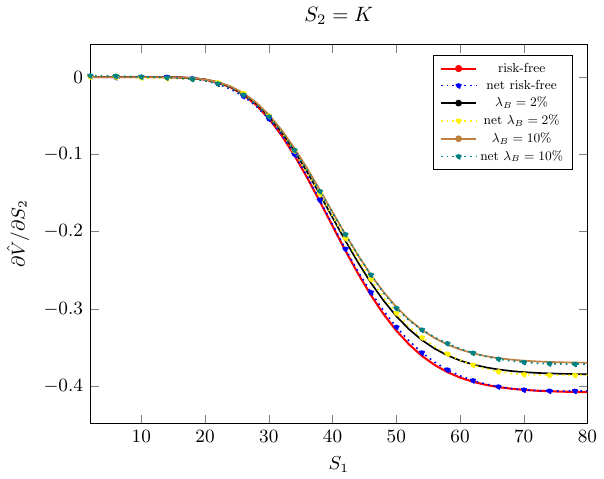}
                \caption{Delta with respect to the $S2$.}
                \label{fig:sellerRatesWorstOfDeltaS2}
            \end{subfigure}
            \caption{Comparison between finite differences and network worst-of put option deltas with $S_2=K$ fixed. Risk-free case ($\lambda_B=\lambda_C=0$), case $\lambda_B=2\%$ and case $\lambda_B=10\%$ are plotted.}
            \label{fig:sellerRatesWorstOfGreeks}
        \end{figure}
        
        In Figure \ref{fig:WorstOfSurface} the network solution for the risky worst-of put option is plotted, while in Figure \ref{fig:WorstOfError} is shown its error compared with the reference solution, both for the case $\lambda_B=2\%$. The maximum relative error remains in the same order of magnitude seen before and present the same pattern as seen before. 
        Figures \ref{fig:sellerRatesWorstOfPrices}, \ref{fig:sellerRatesWorstOfDeltaS1} and \ref{fig:sellerRatesWorstOfDeltaS2} show the comparison between the reference and network approximated prices and first-order derivatives with respect to the $S_1$ and $S_2$, respectively. The results are in line with what we would expect from the arithmetic average option case. In general, it is observed that the behaviour of the derivative is influenced by the direction it follows in relation to the ATM region. Thus, better approximations are obtained when they follow the downward direction, while their quality deteriorates in transverse direction. This situation can be seen in the derivative data given in Table \ref{tab:relErrorWorstOf}, where it can be seen that, if $S_i>S_j,\,i,j=1,2$, then the error in the derivative with respect to the $S_i$ is greater than the error in the derivative with respect to $S_j$, obviously both evaluated at $(S_i, S_j)$. 
        
        \begin{table}[ht]
        \small
           \centering
           \begin{tabular}{c c|c|c|c}
    Case & $(S_1, S_2)$ &  $\hat{V}$ & $\partial \hat{V}/\partial S_1$ & $\partial\hat{V}/\partial S_2$ \\
    \hline
    \multirow{3}{6em}{Risk-free}    & $(50.0, 50.0)$  & $1.34\times 10^{-3}$    & $5.87\times 10^{-3}$    & $1.41\times 10^{-2}$    \\ 
                                    & $(50.0, 45.0)$  & $9.99\times 10^{-4}$    & $1.70\times 10^{-3}$    & $1.22\times 10^{-3}$      \\
                                    & $(45.0, 50.0)$  & $1.02\times 10^{-3}$    & $7.34\times 10^{-3}$    & $1.10\times 10^{-2}$      \\
                                    & $(55.0, 55.0)$  & $5.78\times 10^{-3}$    & $1.56\times 10^{-3}$    & $5.75\times 10^{-4}$    \\
                                    & $(60.0, 53.0)$  & $4.66\times 10^{-3}$    & $1.12\times 10^{-2}$    & $1.06\times 10^{-3}$     \\
                                    & $(53.0, 60.0)$  & $2.57\times 10^{-3}$    & $3.66\times 10^{-4}$    & $5.90\times 10^{-2}$      \\ 
                                    & $(45.0, 45.0)$  & $8.57\times 10^{-4}$    & $1.06\times 10^{-3}$    & $6.16\times 10^{-3}$    \\
                                    & $(57.1, 42.9)$  & $7.42\times 10^{-4}$    & $3.59\times 10^{-3}$    & $1.45\times 10^{-3}$    \\
                                    & $(42.9, 57.1)$  & $1.36\times 10^{-3}$    & $1.79\times 10^{-3}$    & $2.42\times 10^{-2}$      \\

    \hline
    \multirow{3}{6em}{$\lambda_B = 2\%$}    & $(50.0, 50.0)$ & $3.52\times 10^{-3}$    & $1.37\times 10^{-2}$    & $1.24\times 10^{-2}$     \\ 
                                            & $(50.0, 45.0)$ & $8.98\times 10^{-4}$    & $1.00\times 10^{-2}$    & $2.38\times 10^{-3}$     \\ 
                                            & $(45.0, 50.0)$ & $4.48\times 10^{-4}$    & $5.52\times 10^{-3}$    & $1.67\times 10^{-2}$      \\ 
                                            & $(55.0, 55.0)$ & $9.78\times 10^{-3}$    & $2.61\times 10^{-3}$    & $9.30\times 10^{-3}$    \\ 
                                            & $(60.0, 53.0)$ & $9.18\times 10^{-3}$    & $2.30\times 10^{-2}$    & $5.98\times 10^{-3}$      \\ 
                                            & $(53.0, 60.0)$ & $6.59\times 10^{-3}$    & $5.84\times 10^{-3}$    & $6.59\times 10^{-2}$     \\ 
                                            & $(45.0, 45.0)$ & $8.23\times 10^{-4}$    & $9.09\times 10^{-4}$    & $3.68\times 10^{-4}$    \\ 
                                            & $(57.1, 42.9)$ & $2.78\times 10^{-4}$    & $2.09\times 10^{-3}$    & $9.25\times 10^{-3}$     \\ 
                                            & $(42.9, 57.1)$ & $3.16\times 10^{-3}$    & $4.05\times 10^{-3}$    & $2.54\times 10^{-2}$    \\ 
                                        
    \hline
    \multirow{3}{6em}{$\lambda_B = 10\%$}   & $(50.0, 50.0)$ & $3.99\times 10^{-3}$    & $9.38\times 10^{-3}$    & $5.61\times 10^{-3}$      \\ 
                                            & $(50.0, 45.0)$ & $2.04\times 10^{-3}$    & $1.99\times 10^{-4}$    & $2.30\times 10^{-3}$       \\ 
                                            & $(45.0, 50.0)$ & $6.48\times 10^{-4}$    & $3.70\times 10^{-3}$    & $1.31\times 10^{-3}$      \\ 
                                            & $(55.0, 55.0)$ & $5.41\times 10^{-3}$    & $2.18\times 10^{-3}$    & $1.49\times 10^{-2}$       \\ 
                                            & $(60.0, 53.0)$ & $6.75\times 10^{-3}$    & $1.38\times 10^{-2}$    & $1.46\times 10^{-3}$       \\ 
                                            & $(53.0, 60.0)$ & $4.50\times 10^{-3}$    & $1.70\times 10^{-4}$    & $5.26\times 10^{-3}$       \\ 
                                            & $(45.0, 45.0)$ & $1.33\times 10^{-3}$    & $2.96\times 10^{-3}$    & $5.03\times 10^{-3}$      \\ 
                                            & $(57.1, 42.9)$ & $1.23\times 10^{-3}$    & $9.21\times 10^{-3}$    & $3.12\times 10^{-3}$      \\ 
                                            & $(42.9, 57.1)$ & $1.60\times 10^{-3}$    & $1.85\times 10^{-3}$    & $1.43\times 10^{-2}$     \\ 
    \hline

\end{tabular}
           \caption{Relative errors for the worst-of price and deltas. The set of parameters given in Table \ref{tab:BS2d_params} is considered, taking $\lambda_B=\lambda_C=0$ in the risk-free case, and varying $\lambda_B$ in the others. For each case, three ATM, three OTM and three ITM values are taken. }
           \label{tab:relErrorWorstOf}
       \end{table}
   
       
   \subsubsection{Options under the Heston model}
        Finally, we present the results related to the valuation of options using the Heston model, which is based on the description given in Section \ref{sec:Heston}. For this purpose, we work with the model data given in Table \ref{tab:hestonParams}, and we select $S_{max}=4K$, $\nu_{max}=3$, as the truncation values of the domain. 
        
        \begin{table}[!h]
           \centering
           \begin{tabular}{p{7cm}c}
    \hline 
    \multicolumn{2}{c}{Heston parameters}                  \\
    \hline \hline
    Strike, K                               & $1$                  \\
    \hline
    Time to maturity, T                     & $2$                   \\
    \hline
    Repo rate minus dividend, $r_R$         & $0.025$               \\
    \hline
    Interest rate, $r$                      & $0.025$                \\
    \hline
    Mean reversion rate, $\kappa$                    & $1.5$        \\
    \hline
    Mean variance, $\eta$                   &   $0.04$              \\
    \hline
    Volatility of variance, $\sigma$                   &   $0.3$    \\
    \hline
    Correlation, $\rho$                       &   $-0.9$           \\
    \hline \hline
    \multicolumn{2}{c}{xVA parameters}                            \\
    \hline 
    Seller hazard rate, $\lambda_B$         & $[0.0, 0.1]$                \\
    \hline
    Counterparty hazard rate, $\lambda_C$   & $0.04$                \\
    \hline
    Seller recovery rate, $R_B$             & $0.3$                 \\
    \hline
    Counterparty recovery rate, $R_C$       & $0.3$                 \\
    \hline 
    Funding spread, $s_F$                   & $(1 - R_B)\lambda_B$  \\
    \hline
\end{tabular}
           \caption{Parameters for Heston model, adapted from \cite{Hout10adifinite}, and risky parameters.}
           \label{tab:hestonParams}
       \end{table}
        
        We keep the same goal as in the previous cases, namely, the evaluation of the performance of the PINNs algorithm for the Heston's linear and non-linear case. Therefore, we repeat the previously performed experiments considering a put option, so that the loss function is defined by the residuals \eqref{eq:ResHeston_I}-\eqref{eq:resHeston_IC}. We took a total of $349324$ collocation points ($N_T=51, N_S = N_{\nu} = 80$) and set the number of Adam's steps to $25000$, applying the inverse time-decay strategy with $a=10000$ and $\delta=0.5$. The number of L-BFGS steps is remains the same. As in the previous cases, the reference solution is computed with FD, adding a fixed point scheme in the risky cases. 
       
       \begin{figure}[!h]
            \centering
            \begin{subfigure}[b]{0.49\textwidth}
                \centering
                \includegraphics[width=\textwidth]{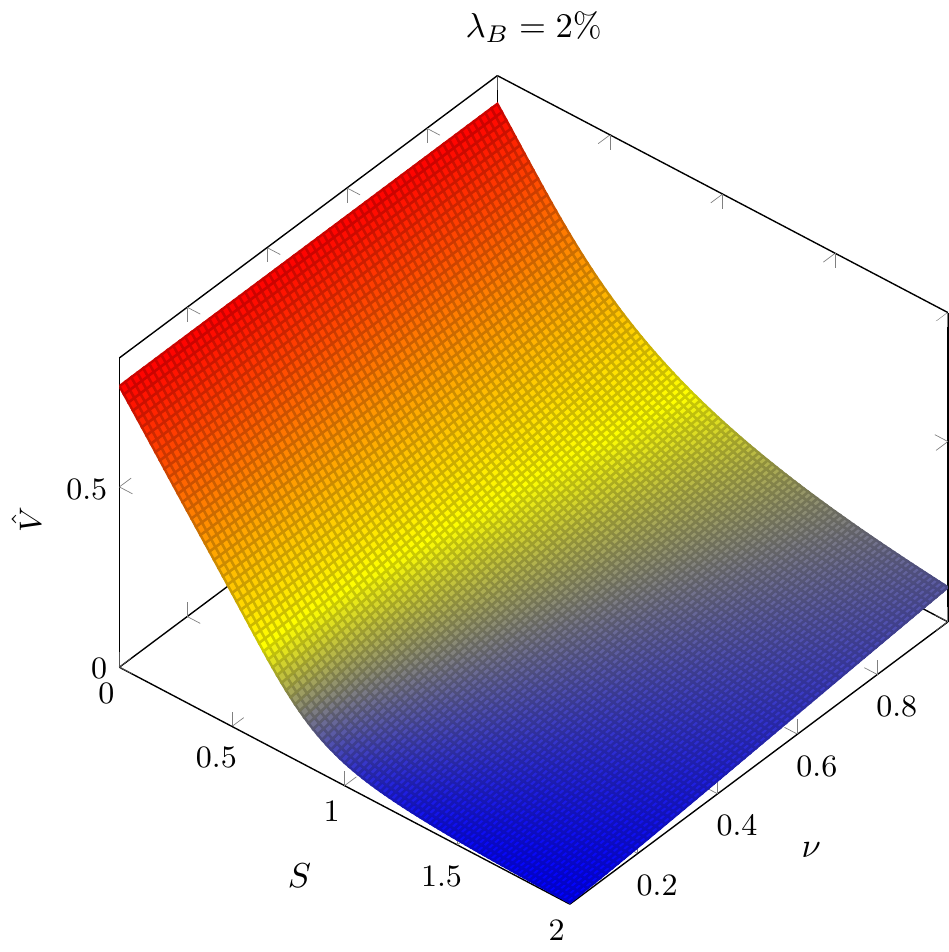}
                \caption{Put option value surface computed via ANN.}
                \label{fig:HestonSurface}
            \end{subfigure}
            \hfill
            \begin{subfigure}[b]{0.49\textwidth}
                \centering
                \includegraphics[width=\textwidth]{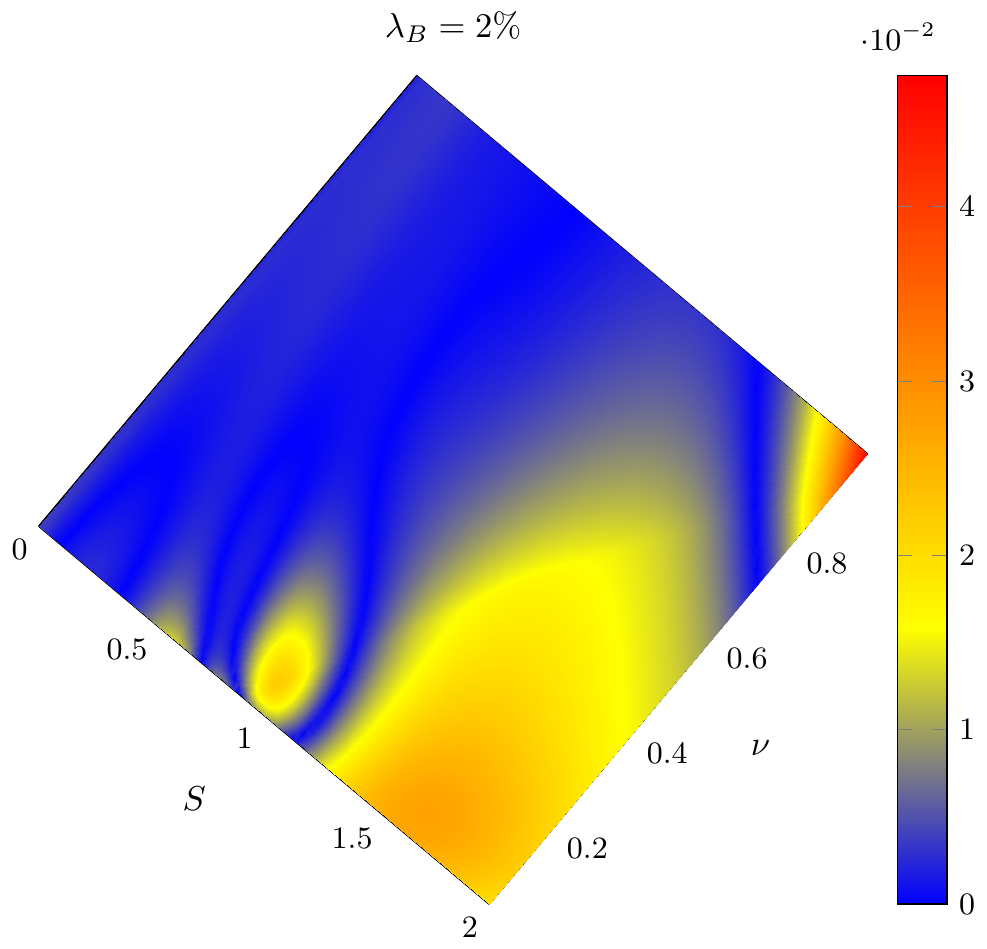}
                \caption{Relative error colour map.}
                \label{fig:HestonError}
            \end{subfigure}
            \caption{Risky put option under the Heston model with parameters given in Table \ref{tab:hestonParams}.}
            \label{fig:HestonPlots}
        \end{figure}
        
        Figure \ref{fig:HestonPlots} shows the price surface computed by the trained ANN (Figure \ref{fig:HestonSurface}), as well as the errors obtained in relation with respect to the reference solution (Figure \ref{fig:HestonError}), for the case $\lambda_B=2\%$. Both the qualitative and quantitative behaviour of the solution achieve the precision standards of the other two-dimensional cases studied above. However, a different distribution of the committed error is observed. In previous cases, the error was concentrated in the ATM region, mostly due to the non-differentiability of the payoff. Now, although we see the expected larger error in the ATM region when the values of $\nu$ are close to zero, it becomes dominant in the OTM region. Such error pattern has also been found in FD algorithms. This fact suggests that the chosen boundary conditions due to the truncation domain could  be hampering the accuracy of the approximation. 
        
\begin{figure}[!h]
            \centering
            \begin{subfigure}[b]{0.49\textwidth}
                \centering
                \includegraphics[width=\textwidth, page=1]{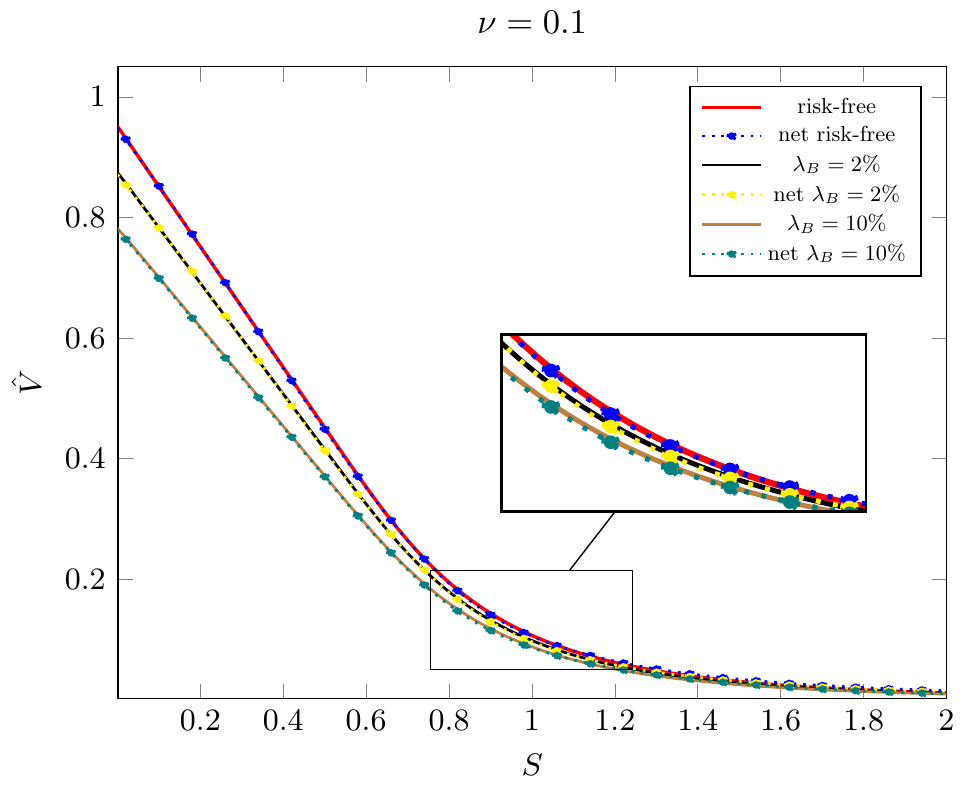}
            \end{subfigure}
            \hfill
            \begin{subfigure}[b]{0.49\textwidth}
                \centering
                \includegraphics[width=\textwidth, page=2]{images/2D/PLOTS_SellerRatesHeston.pdf}
            \end{subfigure}
            \caption{Comparison between finite differences and network put option under the Heston model. Risk-free case ($\lambda_B=\lambda_C=0$), case $\lambda_B=2\%$ and case $\lambda_B=10\%$ are plotted.}
            \label{fig:sellerRatesWHestonPrices}
        \end{figure}

\begin{figure}[!h]
            \centering
            \begin{subfigure}[b]{0.49\textwidth}
                \centering
                \includegraphics[width=\textwidth, page=3]{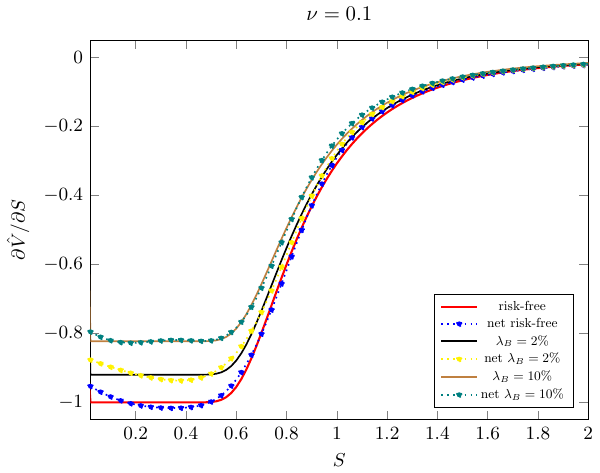}
            \end{subfigure}
            \hfill
            \begin{subfigure}[b]{0.49\textwidth}
                \centering
                \includegraphics[width=\textwidth, page=4]{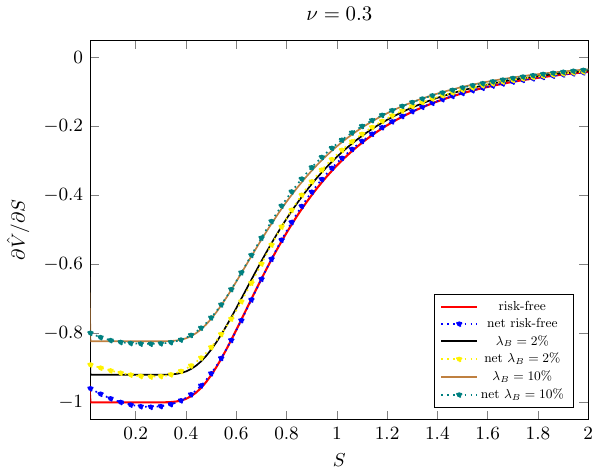}
            \end{subfigure}
            \caption{Comparison between finite differences and network put option deltas under the Heston model. Risk-free case ($\lambda_B=\lambda_C=0$), case $\lambda_B=2\%$ and case $\lambda_B=10\%$ are plotted.}
            \label{fig:sellerRatesWHestonDeltas}
        \end{figure}

\begin{figure}[!h]
            \centering
            \begin{subfigure}[b]{0.49\textwidth}
                \centering
                \includegraphics[width=\textwidth, page=5]{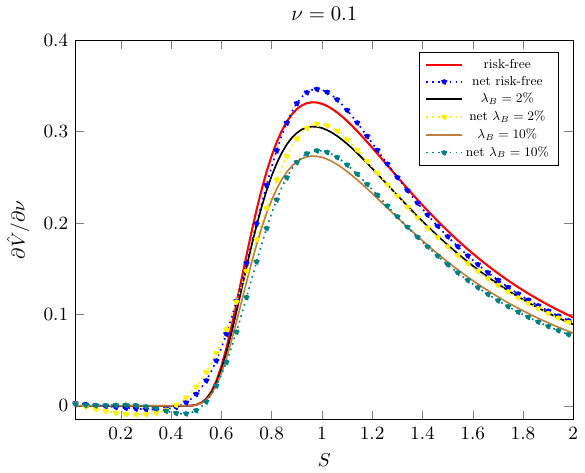}
            \end{subfigure}
            \hfill
            \begin{subfigure}[b]{0.49\textwidth}
                \centering
                \includegraphics[width=\textwidth, page=6]{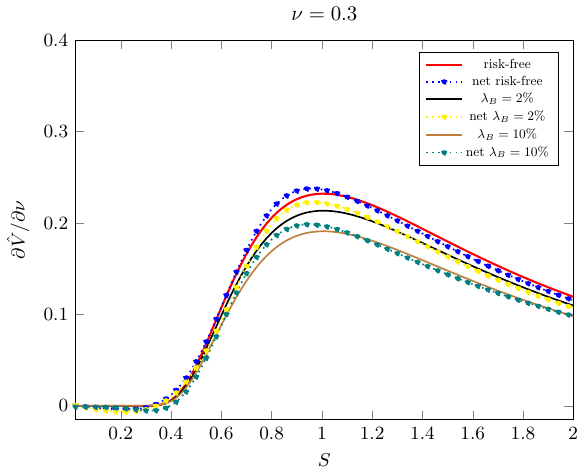}
            \end{subfigure}
            \caption{Comparison between finite differences and network put option vegas under the Heston model. Risk-free case ($\lambda_B=\lambda_C=0$), case $\lambda_B=2\%$ and case $\lambda_B=10\%$ are plotted.}
            \label{fig:sellerRatesWHestonVegas}
        \end{figure}

        In the Figures \ref{fig:sellerRatesWHestonPrices}, \ref{fig:sellerRatesWHestonDeltas} and \ref{fig:sellerRatesWHestonVegas}, $\nu$-slices of the solution and its first order derivatives are shown. Such slices correspond to sections with $\nu=0.1$ and $\nu=0.3$ (values of interest in the industry). As in the examples seen above, the risk-free case and the cases with $\lambda_B = 2\%, 10\%$ are considered. In the price plots (Figure \ref{fig:sellerRatesWHestonPrices}), a similar performance to that seen in the previous cases can be observed. This results are supported by the Table \ref{tab:hestonErrors}, where the relative errors obtained for the points of interest are shown, achieving, at least, an order of $10^{-2}$.
        
        For the first-order derivatives, the obtained accuracy is sufficient for financial purposes, although a slight decrease in the performance is found due to the more complex physics described by the PDE. In the case of deltas, Figure \ref{fig:sellerRatesWHestonDeltas}, the oscillatory behaviour near the $S=0$ boundary seen before is slightly magnified, specially for the risk-free and lower $\lambda_B$ scenarios. However, it is able to perfectly capture its asymptotic behaviour as $S$ grows. The approximations around the strike are remarkably good, with relative errors an the order of $10^{-2}$, see Table \ref{tab:hestonErrors}. Figure \ref{fig:sellerRatesWHestonVegas} shows the vega slices, understanding vega as the derivative of the price  with respect to the underlying's variance\footnote{We assume an abuse of language. In reality, vega is understood as the partial derivative with respect to the square root of the variance but, considering fixed-$\nu$ slices, both expressions only differ in being multiplied by a constant.}. Regardless of the chosen default scenario,  the approximations close to $S=0$ are worse. Moreover, the estimations are affected by the closeness to the boundary $\nu=0$, so that the closer you are to such boundary, the lower the accuracy is. However, this effect looses intensity or directly disappears for larger values of $S$. Thus, the same order as that obtained for the deltas is observed in the neighbourhood of the strike, see Table \ref{tab:hestonErrors}, and the asymptotic behaviour is consistent with the Neumann condition imposed on the $\nu_{max}$ boundary.

        \begin{table}[!h]
            \centering
                \begin{tabular}{c c|c|c|c}
    Case & $(S, \nu)$ &  $\hat{V}$ & $\partial \hat{V}/\partial S$ & $\partial\hat{V}/\partial \nu$ \\
    \hline
    \multirow{3}{6em}{Risk-free}    & $(0.8, 0.1)$  & $7.82\times 10^{-3}$    & $2.76\times 10^{-2}$    & $4.46\times 10^{-2}$   \\ 
                                    & $(0.8, 0.3)$  & $2.33\times 10^{-3}$    & $1.05\times 10^{-2}$    & $4.57\times 10^{-2}$   \\ 
                                    & $(1.0, 0.1)$  & $1.18\times 10^{-2}$    & $5.62\times 10^{-2}$    & $4.58\times 10^{-2}$   \\ 
                                    & $(1.0, 0.3)$  & $6.59\times 10^{-3}$    & $2.37\times 10^{-2}$    & $1.97\times 10^{-2}$   \\ 
                                    & $(1.2, 0.1)$  & $3.33\times 10^{-2}$    & $7.48\times 10^{-2}$    & $1.80\times 10^{-2}$   \\ 
                                    & $(1.2, 0.3)$  & $1.81\times 10^{-2}$    & $1.21\times 10^{-2}$    & $1.44\times 10^{-2}$   \\ 
  
    \hline
    \multirow{3}{6em}{$\lambda_B = 2\%$}   & $(0.8, 0.1)$ & $5.95\times 10^{-3}$    & $3.65\times 10^{-2}$    & $7.60\times 10^{-2}$   \\ 
                                            & $(0.8, 0.3)$ & $4.61\times 10^{-3}$    & $3.24\times 10^{-3}$    & $5.02\times 10^{-2}$   \\ 
                                            & $(1.0, 0.1)$ & $2.02\times 10^{-2}$    & $4.03\times 10^{-2}$    & $1.43\times 10^{-2}$   \\ 
                                            & $(1.0, 0.3)$ & $1.12\times 10^{-3}$    & $2.28\times 10^{-2}$    & $4.02\times 10^{-2}$   \\ 
                                            & $(1.2, 0.1)$ & $1.11\times 10^{-2}$    & $7.40\times 10^{-2}$    & $5.87\times 10^{-3}$   \\ 
                                            & $(1.2, 0.3)$ & $1.21\times 10^{-2}$    & $2.39\times 10^{-2}$    & $1.07\times 10^{-2}$   \\

    \hline
    \multirow{3}{6em}{$\lambda_B = 10\%$}   & $(0.8, 0.1)$ & $1.81\times 10^{-2}$    & $1.72\times 10^{-2}$    & $6.38\times 10^{-2}$   \\ 
                                            & $(0.8, 0.3)$ & $7.37\times 10^{-3}$    & $2.14\times 10^{-2}$    & $7.56\times 10^{-2}$   \\ 
                                            & $(1.0, 0.1)$ & $2.05\times 10^{-2}$    & $5.99\times 10^{-2}$    & $2.59\times 10^{-2}$   \\ 
                                            & $(1.0, 0.3)$ & $2.48\times 10^{-3}$    & $2.76\times 10^{-2}$    & $3.25\times 10^{-2}$   \\ 
                                            & $(1.2, 0.1)$ & $1.87\times 10^{-2}$    & $7.35\times 10^{-2}$    & $1.85\times 10^{-2}$   \\ 
                                            & $(1.2, 0.3)$ & $1.29\times 10^{-2}$    & $9.79\times 10^{-3}$    & $1.08\times 10^{-2}$   \\

    \hline
\end{tabular}

            \caption{Relative errors for the put option price and first order derivatives, under the Heston model. The set of parameters given in Table \ref{tab:hestonParams} is considered, taking $\lambda_B=\lambda_C=0$ in the risk-free case, and varying $\lambda_B$ in the others. We consider two variance for the ITM, ATM and OTM case.}
            \label{tab:hestonErrors}
        \end{table}
    
    \clearpage
    \section{Conclusions} \label{sec:conclusions}
    
Thanks to the universal approximation property of ANNs and the dramatic increase of computing power of deep learning hardware, PINNs methods have become a serious alternative for solving hard PDE problems. Maybe the biggest weakness of PINNs is the imposition of the boundary conditions, as they enter as addends into the loss function for the network calibration, and the user must choose heuristically the magnitude of the addends that depend, of course, on the type of problem and the type of boundary conditions. These weights are not known a priori, as they depend of the solution itself, and must be estimated in some way, which is also problem dependent.

In this work a novel technique for the treatment of the boundary conditions in the PINNs framework has been introduced. It allows to get ride the heuristic selection of the weights of the boundary addends that appear in the loss function of the ANN that approximates the solution. The strategy is based on the direct evaluation of the differential operator at the boundaries taking into account the imposed boundary conditions. This yields an addend for the boundary that is in the same magnitude of the loss in the interior of the domain, avoiding to deal with the heuristic choice of the weights. To the best of our knowledge this procedure is introduced in this paper for the first time, and we feel that is a very interesting contribution that makes PINNs much more powerful and easier to use.  

The new approach has been applied to several non-linear PDE problems that arise in computational finance when CCR is taking into account, although it is general enough and non problem-dependent to be applied in other fields, like for example fluid dynamics or solid mechanics. In particular, it has been employed to solve the boundary value problems related to the pricing of risky European options under the one and two-dimensional Black-Scholes model, as well as under the Heston model. The obtained solutions yield a good accuracy when compared with the analytical or reference solutions. Furthermore, embedding the obtained solution into an ANN has allowed us to compute their relevant partial derivatives by means of AD. 

All in all, the partial derivatives' computation in the PINNs framework is so far a generally unexplored avenue and we believe it may have a lot of potential, being one of the main advantages of PINNs over other deep-learning based methodologies. The optimization procedure takes these quantities into account since they are implicitly part of the loss function, so that under the assumption of having an ideal optimizer, there would be a perfect fit of both the solution and the derivatives that conform the PDE. Another of the most notables advantages is in terms of interpretability. Compared to other techniques, this methodology is closer to the classical PDE schemes, in the sense that the PDE solution is projected onto a space formed by the ANN weights.

    \section*{Acknowledgements}
    All the authors thank to the support received from the CITIC research center, funded by Xunta de Galicia
and the European Union (European Regional Development Fund - Galicia Program), by grant ED431G 2019/01.

A.L and J.A.G.R. acknowledge the support received by the Spanish MINECO under research project number PDI2019-108584RB-I00, and by the Xunta de Galicia under grant ED431C 2018/33.

    \clearpage
    \newpage
    
    \bibliographystyle{plain}
    \bibliography{references.bib}

\end{document}